 \documentclass[final,3p,times]{elsarticle}

\usepackage{amssymb}
\usepackage{babel,tabularx,ragged2e,booktabs}

\usepackage{lineno}

\usepackage[utf8]{inputenc}
\usepackage{textcomp}
\usepackage[mathscr]{euscript}
\usepackage{graphicx}
\usepackage{amsmath}
\usepackage[version=4]{mhchem}
\usepackage{siunitx}
\usepackage{longtable,tabularx}
\usepackage{color}
\usepackage{cancel}
\usepackage{gensymb}
\usepackage{color}
\usepackage{tcolorbox}
\usepackage{float}
\usepackage{multirow}
\usepackage{caption}
\usepackage{subcaption}
\usepackage{graphicx}
\usepackage{amsmath}
\usepackage{multicol}
\usepackage{wrapfig}
\usepackage{amsmath,amssymb}
\usepackage{nicematrix}
\usepackage{paracol}
\usepackage{ragged2e}
\usepackage{algpseudocode}
\usepackage{algorithm2e}
\RestyleAlgo{ruled}

\usepackage{tikz}

\makeatletter
\newsavebox\myboxA
\newsavebox\myboxB
\newlength\mylenA

\newcommand*\xoverline[2][0.75]{%
    \sbox{\myboxA}{$\m@th#2$}%
    \setbox\myboxB\null
    \ht\myboxB=\ht\myboxA%
    \dp\myboxB=\dp\myboxA%
    \wd\myboxB=#1\wd\myboxA
    \sbox\myboxB{$\m@th\overline{\copy\myboxB}$}
    \setlength\mylenA{\the\wd\myboxA}
    \addtolength\mylenA{-\the\wd\myboxB}%
    \ifdim\wd\myboxB<\wd\myboxA%
       \rlap{\hskip 0.5\mylenA\usebox\myboxB}{\usebox\myboxA}%
    \else
        \hskip -0.5\mylenA\rlap{\usebox\myboxA}{\hskip 0.5\mylenA\usebox\myboxB}%
    \fi}
\makeatother

\usepackage[colorlinks]{hyperref}

\colorlet{colorRev1}{blue!80!black} 
\colorlet{colorRev2}{red!80!black} 

\journal{Aerospace Science and Technology}

\begin{document}

\begin{frontmatter}



\title{Dynamics modelling and path optimization for the on-orbit assembly of large flexible structures using a multi-arm robot}


\author[ISAE]{R. Rodrigues\corref{cor1}}
\ead{ricardo.rodrigues@isae-supaero.fr}
\cortext[cor1]{Corresponding author; Ph.D. student.}

\author[ESA]{V. Preda}
\ead{valentin.preda@esa.int}

\author[ISAE]{F. Sanfedino\fnref{label2}}
\ead{francesco.sanfedino@isae-supaero.fr}
\fntext[label2]{Associate Professor}

\author[ISAE]{D. Alazard\fnref{label3}}
\ead{daniel.alazard@isae-supaero.fr}
\fntext[label3]{Professor}

\address[ISAE]{ Fédération ENAC ISAE-SUPAERO ONERA, Université de
Toulouse, 10 Avenue Edouard Belin, BP-54032, 31055, Toulouse, France}
\address[ESA]{European Space Agency (ESA)/ESTEC, Keplerlaan 1, 2201 AZ, Noordwijk, The Netherlands}

\begin{abstract} 

This paper presents a comprehensive methodology for modeling an on-orbit assembly mission scenario of a large flexible structure using a multi-arm robot. This methodology accounts for significant changes in inertia and flexibility throughout the mission, addressing the problem of coupling dynamics between the robot and the evolving flexible structure during the assembly phase. A three-legged walking robot is responsible for building the structure, with its primary goal being to walk stably on the flexible structure while picking up, carrying and assembling substructure components. To accurately capture the dynamics and interactions of all subsystems in the assembly scenario, various linear fractional representations (LFR) are developed, considering the changing geometrical configuration of the multi-arm robot, the varying flexible dynamics and uncertainties. A path optimization algorithm is proposed for the multi-arm robot, capable of selecting trajectories based on various cost functions related to different performance and stability metrics. The obtained results demonstrate the effectiveness of the proposed modeling methodology and path optimization algorithm.

\end{abstract}


\end{frontmatter}


\section*{Nomenclature}

{

\noindent\begin{longtable*}{@{}l @{\quad \quad} l@{}}
{AOCS} & {Attitude and Orbital Control System} \\
{DCM} & {Direct Cosine Matrix} \\
{DOF} & {Degree Of Freedom} \\
{ESA} & {European Space Agency} \\
{FEM} & {Finite Element Model} \\
{GNC} & {Guidance, Navigation and Control} \\
{JWST} & {James Webb Space Telescope} \\
{LFR} & {Linear Fractional Representation} \\
{LPV} & {Linear Parameter-Varying} \\
{MAR} & {Multi-Arm Robot} \\
{SBSP} & {Space-Based Solar Power} \\
{SDTlib} & {Satellite Dynamics Toolbox library} \\

{TITOP} & {Two-Input Two-Output Ports} \\
{3D} & {Three-Dimensional} \\

$\displaystyle \mathbf{a}_{P}$ & Inertial linear acceleration vector at the point $P$, expressed in \si{\meter\per\square\second}.\\

$\boldsymbol{\dot{\omega}}_P$ & Inertial angular acceleration vector at the point $P$, expressed in \si{\radian\per\square\second}.\\

\vspace{1mm}
$\boldsymbol{{\omega}}_P$ & Inertial angular velocity vector at the point $P$, expressed in \si{\radian\per\second}. \\
$ \mathbf{OP}$ & Distance vector defined between the point $O$, the origin of the inertial frame, \\
& and the point ${P}$, expressed in \si{\meter}. \\

$\boldsymbol{{\Theta}}_{P}$ & \textsc{Euler} angles vector computed at the point $P$ with respect to the inertial \\
& frame, expressed in \si{\radian} and using the 'ZYX' sequence of rotations. \\
$\mathbf{F}_{\mathrm{ext} / \mathcal{B}, G}$  & External forces vector applied to the body $\mathcal{B}$ at the point $G$, expressed in \si{\newton}.\\
$\mathbf{T}_{\mathrm{ext} / \mathcal{B}, G}$ & External torques vector applied to the body 
$\mathcal{B}$ at the point $G$, expressed in \si{\newton\meter}.\\
$\mathbf{F}_{\mathcal{B/A},G}$ & Forces vector applied by the body 
$\mathcal{B}$ to the body $\mathcal{A}$ at the point $G$, expressed in \si{\newton}.\\
$\mathbf{T}_{\mathcal{B/A},G}$ & Torques vector applied by the body 
$\mathcal{B}$ to the body $\mathcal{A}$ at the point $G$, expressed in 
\si{\newton\meter}.\\
$\mathbf{W}_{\mathcal{B/A},G}$ & Wrench vector applied by the body $\mathcal{B}$ to the body $\mathcal{A}$ at the point $G$:\\
& $\mathbf{W}_{\mathcal{B/A},G}=\left[\mathbf{F}^{\mathrm{T}}_{\mathcal{B/A},G}, \mathbf{T}^{\mathrm{T}}_{\mathcal{B/A},G}\right]^{\mathrm{T}}$, expressed in $\left[\si{\newton}, \si{\newton\meter}\right]$.\\

$\left[\mathbf{X}\right]_{\mathcal{R}_{\bullet}}$ & $\mathbf{X}$ (model, vector 
or tensor) projected in the frame $\mathcal{R}_{\bullet}$.\\
$\mathbf{X}$ & $\mathbf{X}$ (model, vector 
or tensor) projected in the body frame, unless stated otherwise.\\

$\mathbf P_{\mathcal R_a/\mathcal R_b}$ & DCM from the frame $\mathcal R_a$ to the frame $\mathcal R_b$ ($\left[\mathbf{v}\right]_{\mathcal{R}_{b}}=\mathbf P_{\mathcal R_a/\mathcal R_b}\left[\mathbf{v}\right]_{\mathcal{R}_{a}}$ for any vector $\mathbf v$).\\

$\textbf{I}_{n}$ &  Identity matrix $n \times n$.\\
$\mathbf{0}_{n\times m}$  & Zero matrix $n \times m$.\\
$\mathbf{v}\{i\}$ & Component $i$ of vector $\mathbf{v}$.\\
$(^*\mathbf{v})$ & Skew symmetric matrix associated with vector
$\mathbf{v}$: $\left[(^*\mathbf{v})\right]_{\mathcal{R}_{\bullet}}=\left[\begin{array}{ccc}0 & -\mathbf{v}\{3\} & \mathbf{v}\{2\} \\ \mathbf{v}\{3\} & 0 & -\mathbf{v}\{1\} \\ -\mathbf{v}\{2\} & \mathbf{v}\{1\} & 0 \end{array}\right]_{\mathcal{R}_{\bullet}}$.\\
$\boldsymbol{\tau}_{PB}$ &  Kinematic model between the points $P$ and $B$: $\left[\boldsymbol{\tau}_{PB}\right]_{\mathcal{R}_{\bullet}}=\left[\begin{array}{cc} \textbf{I}_{3}& (^*\mathbf{PB}) \\  \mathbf{0}_{3\times 3}& \textbf{I}_{3} \end{array}\right]_{\mathcal{R}_{\bullet}}$.\\

$\mathbf{X}_{(\mathbf{I},\mathbf{J})}$ & Subsystem of $\mathbf{X}$ from the inputs indexed in the vector $\mathbf{J}$ to the outputs indexed \\ 
& in the vector $\mathbf{I}$ (if $(\mathbf{I},\mathbf{J})=(:,7:10)$, it means that one is considering the subsystem \\ 
& between the inputs from $7$ until $10$ to all the outputs). \\
\vspace{1mm}
$\textbf{J}_{G}^{\mathcal{B}}$ & Inertia tensor of $\mathcal{B}$, computed at the point $G$, written in $\mathcal{R}_{b}$ and expressed in \si{\kilogram\square\meter}.\\
$m^{\mathcal{B}}$ & Mass of $\mathcal{B}$ expressed in \si{\kilogram}.\\
$\mathrm{s}$ & \textsc{Laplace}'s variable.

\end{longtable*}}

\section{Introduction}

\subsection{Background and motivation}

The dimensions of large flexible structures in orbit, such as telescopes, solar arrays, mirrors or antennas \cite{antennaGMV}, are currently restricted by the need to launch these structures as a single piece within the maximum allowable volume of launch vehicle fairings \cite{pirat2022}. For example, the James Webb Space Telescope (JWST) was designed to be foldable to fit these constraints, involving a complex folding process with multiple deployable structures and release mechanisms. Other examples of deployable structures for space missions are also discussed in the literature \cite{santiago2013,PUIG201012, MA2022207}.

The space-based solar power (SBSP) \cite{YANG2023108406} concept has also been extensively studied in the past, with recent renewed interest \cite{urbina2023}. To prepare for future decision-making on SBSP, the European Space Agency (ESA) has launched a preparatory initiative called SOLARIS \cite{kulu2023}. Previous studies have highlighted the challenges of implementing SBSP due to the need for extremely large infrastructures and numerous heavy launches. Given the constraints of a launch vehicle's carrying capacity, constructing these large space structures requires either on-orbit deployment or assembly \cite{wang2021}. Recently, numerous researchers have indicated that on-orbit assembly is the most viable approach to address this issue \cite{she2019}, with support from advancements in space robotics and modular structure designs. One common approach for building such structures is to rely on robotic manipulators \cite{Nanjangud,Lee,Oegerle, jenett}, meaning that on-orbit assembly using robotic systems in weightlessness is expected to be essential in the future. Various enabling technologies could be utilized, including the HOTDOCK \cite{hotdock} modular interconnect, the MOSAR walking manipulator \cite{mosar} and the ESA's Multi-Arm Robot \cite{mirrormar} (MAR).

Few studies have focused on the modeling and control of on-orbit assembly mission scenarios. Wang \textit{et al.} worked on a distributed adaptive vibration control methodology for solar power satellites during on-orbit assembly \cite{WANG2019105378}, whereas Dong \textit{et al.} investigated the dynamic modeling and attitude control for large modular antennas in on-orbit assembly \cite{DONG2024108959}. Additionally, Zhang \textit{et al.} described a modular robotic manipulator that offers high precision and flexibility \cite{zhang2024}, while Zhou \textit{et al.} proposed a dynamic model that accurately describes the dynamic characteristics of large-scale space structures during on-orbit assembly \cite{Zhou2023}. However, no studies in the literature have proposed path optimization for a multi-arm robot assembling a flexible structure on-orbit to minimize specific performance and stability metrics.

From the perspective of the Attitude and Orbit Control System (AOCS) and Guidance, Navigation, and Control (GNC), the challenges associated with on-orbit assembly missions using a multi-arm robot arise primarily due to the highly time-varying flexible dynamics and inertial properties, as well as the dynamic couplings between the moving robot and the flexible structure. Large flexible space structures often have distinctive characteristics, including low resonant frequencies, very low natural damping and stringent mission requirements, such as high pointing accuracy and precise alignment/orientation \cite{Sean2020}. Consequently, when such a structure is disturbed or excited, it tends to remain in vibration for an extended period of time \cite{Gong2019,Peggy}. Additionally, the coupling dynamics between the robot and the structure pose a unique challenge for in-space robotic assembly, as the interaction can induce structural vibrations and degrade the performance of the assembly process. Therefore, the success of such missions relies heavily on the ability to develop an accurate system model \cite{yoshida,maghami}. In this context, the comprehensive modeling of complex multibody structures becomes crucial as it enables the prediction of worst-case scenarios early on.

One of the main contributions of this paper is to propose a methodology for modeling the linear dynamic behavior of an on-orbit assembly scenario of a large flexible structure by means of a three-legged robot. The proposed model is built using the Two-Input-Two-Output Port (TITOP) approach \cite{alazard}, which considers forces and accelerations at the connection points as inputs and outputs. Unlike traditional methods, this approach is not reliant on specific boundary conditions at the connection points of the body being modeled. This approach offers the possibility to model complex multibody mechanical systems, while keeping the uncertain nature of the plant and condensing all the possible mechanical configurations in Linear Fractional Representation (LFR) sytems. This approach has been previously introduced in \cite{alazard} and has found practical applications in space engineering, as demonstrated in \cite{MURALI, Perez2015, Perez2016, SANFEDINO2018128, RODRIGUES2022107865, rodrigues, Rodrigues2024ModelingAA}. The obtained LFR models are characterized by their parameterization with respect to the geometrical configuration of the moving robot, resulting in  Linear Parameter-Varying (LPV) systems that completely accommodate the time varying inertial and flexible dynamics.

The models constructed using the TITOP approach are well-suited for robust control synthesis, as well as robust performance assessment. Additionally, all the models developed with the TITOP approach have been integrated into the most recent release of the Satellite Dynamics Toolbox library (SDTlib) \cite{userguide}.

The assembly of flexible structures demands a precision and adaptability that multi-arm installation robots are particularly suited to provide. As technology progresses, these robots will become increasingly vital in unlocking the potential of on-orbit assembly for the space industry. Therefore, the second main objective of this paper is to develop an innovative method for optimizing the path of a multi-arm robot when installing modular tiles and moving across the flexible structure, while complying with various performance and stability requirements which are imposed on the different models describing the entire assembly scenario. This framework is designed to offer insights into the dynamics of the assembly process and can be used to generate trajectories for constructing any large and flexible space structure.

For the on-orbit assembly mission scenario being studied in this paper, a single spacecraft is considered. This spacecraft consists of a rigid central body, a flexible solar array, a multi-arm installation robot with three robotic arms and a stack of modular tiles ready for assembly. The multi-arm robot is tasked with picking up these tiles one by one and assembling a large flexible structure attached to the spacecraft. It should be noted that the chosen structure type is merely to demonstrate the proposed approach's capabilities. The same toolset can be used to explore the assembly of other structures.

\subsection{Contributions and paper organization}

The paper introduces the following key contributions:

\begin{itemize}

\item the development of several models fully capturing the dynamics and interactions between all subsystems of an on-orbit assembly scenario: moving multi-arm robot and time-varying flexible appendages. These models include the various interactions and parameterization effects in a very compact representation. 

\item a new method for optimizing the path of a multi-arm robot during the on-orbit assembly of a flexible structure. This algorithm incorporates various cost functions, each defined by different performance and stability metrics that are imposed on the different models representing the dynamics of the complete mission scenario.

\end{itemize}

This paper is divided into two main sections: system modeling and path optimization. In the first section (section \ref{modeling}), the TITOP approach is introduced and several linear fractional representations are developed, fully describing the mission's time-varying inertial and flexible properties, as well as the dynamic couplings. In the second section (section \ref{pathopt}), a comprehensive explanation of the path optimization algorithm is provided. This algorithm integrates node graph theory with the linear models that accurately represent the assembly system, optimizing the multi-arm trajectories for assembling a large flexible structure in orbit. The optimization process aims to minimize various performance and stability requirements imposed as constraints on the loop. Finally, results are presented to demonstrate the effectiveness of the proposed method.

\section{Multibody Modeling Approach}
\label{modeling}

\begin{figure}[!ht]
\centering
 \includegraphics[width=1\textwidth]{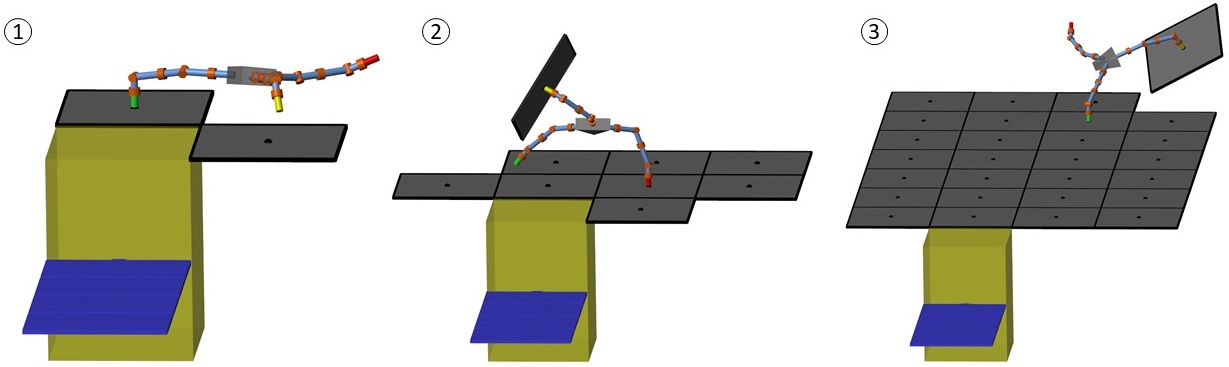}
\caption{Three different illustrations of the on-orbit assembly mission scenario being studied: {\normalsize \textcircled{\scriptsize 1}} the green robotic arm is attached to the initial and sole assembled tile, while the yellow robotic arm is extending to pick another tile from the tile stack;  {\normalsize \textcircled{\scriptsize 2}} the multi-arm robot is moving across the structure while transporting the eighth tile for assembly; {\normalsize \textcircled{\scriptsize 3}} the final tile from the stack has been picked up by the multi-arm robot, which is now adding it to the rest of the assembled structure.}
\label{Diapositive1} 
\end{figure}

For the on-orbit assembly mission scenario being studied in this paper, one spacecraft is considered. This spacecraft is composed of one rigid central body, one flexible solar array, one multi-arm installation robot with three robotic arms and a stack of modular tiles ready to be assembled. The multi-arm robot is responsible for picking up these tiles one by one and assembling a big flexible structure that is clamped to the spacecraft. For a better understanding of the mission scenario being studied, Fig. \ref{Diapositive1} depicts three different representations of the system during the whole assembly phase. 

\subsection{The TITOP approach}

\begin{figure}[!ht]
\centering
 \includegraphics[width=1\textwidth]{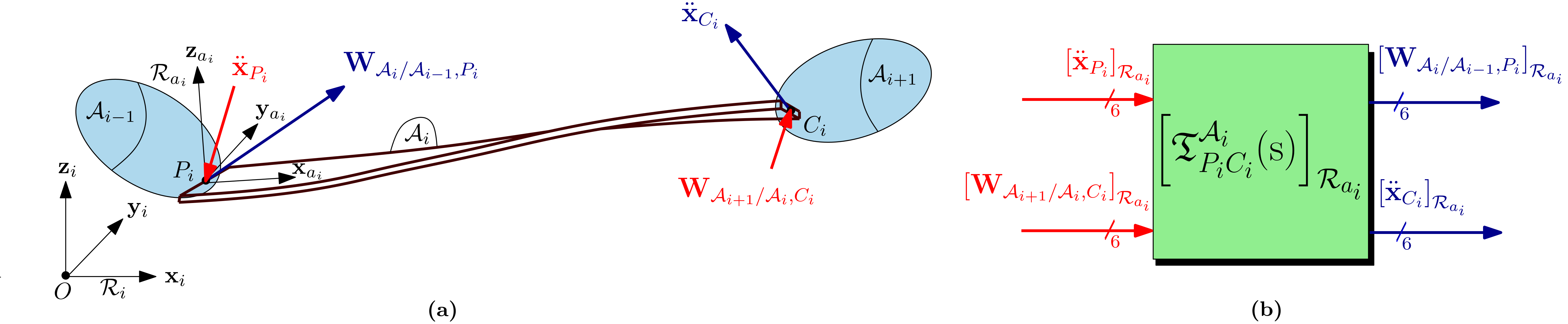}
\caption{(a) TITOP illustration for a generic flexible body $\mathcal{A}_i$. (b) TITOP model $\left[\mathfrak{T}_{P_{i}C_{i}}^{\mathcal{A}_{i}}(\mathrm s)\right]_{\mathcal{R}_{a_i}}$ block-diagram.}
\label{TITOP} 
\end{figure}

The flexible body $\mathcal{A}_{i}$ connected to the parent substructure $\mathcal{A}_{i-1}$ at point $P_{i}$ and to the child substructure $\mathcal{A}_{i+1}$ at point $C_{i}$ is depicted in Fig. \ref{TITOP}a. The double-port or TITOP model ${\left[\mathfrak{T}_{P_{i} C_{i}}^{\mathcal{A}_{i}}(\mathrm{s})\right]}_{\mathcal{R}_{a_i}}$ is a linear dynamic model between 12 inputs:
\begin{itemize}
    
\item the six components of the wrench ${\left[\mathbf{W}_{\mathcal{A}_{i+1}/\mathcal{A}_{i},{C_{i}}}\right]}_{\mathcal{R}_{a_i}}=\left[\begin{array}{c} \mathbf{F}_{\mathcal{A}_{i+1}/\mathcal{A}_{i},{C_{i}}} \\ \mathbf{T}_{\mathcal{A}_{i+1}/\mathcal{A}_{i},{C_{i}}} \end{array}\right]_{\mathcal{R}_{a_i}}$ applied by the child substructure $\mathcal{A}_{i+1}$ to the body $\mathcal{A}_{i}$ at point $C_{i}$, expressed in the body frame  $\mathcal{R}_{a_i}=\left(P_i; \mathbf{x}_{a_i}, \mathbf{y}_{a_i}, \mathbf{z}_{a_i}\right)$.
\item the six components of the acceleration twist ${\left[\ddot{\mathbf{x}}_{P_{i}}\right]}_{\mathcal{R}_{a_i}}=\left[\begin{array}{c} \mathbf{a}_{P_{i}} \\ \boldsymbol{\dot{\omega}}_{P_{i}}\end{array}\right]_{\mathcal{R}_{a_i}}$, expressed in the body frame $\mathcal{R}_{a_i}$.
\end{itemize}
and 12 outputs:
\begin{itemize}
\item the six components of the acceleration twist ${\left[\ddot{\mathbf{x}}_{C_{i}}\right]}_{\mathcal{R}_{a_i}}=\left[\begin{array}{c} \mathbf{a}_{C_{i}} \\ \boldsymbol{\dot{\omega}}_{C_{i}}\end{array}\right]_{\mathcal{R}_{a_i}}$, expressed in the body frame $\mathcal{R}_{a_i}$.
\item the six components of the wrench ${\left[\mathbf{W}_{\mathcal{A}_i/\mathcal{A}_{i-1},{P_{i}}}\right]}_{\mathcal{R}_{a_i}}=\left[\begin{array}{c} \mathbf{F}_{\mathcal{A}_i/\mathcal{A}_{i-1},{P_{i}}} \\ \mathbf{T}_{\mathcal{A}_i/\mathcal{A}_{i-1},{P_{i}}} \end{array}\right]_{\mathcal{R}_{a_i}}$ that is applied by the body $\mathcal{A}_{i}$ to the parent substructure $\mathcal{A}_{i-1}$ at point $P_{i}$, expressed in the body frame $\mathcal{R}_{a_i}$.
\end{itemize}

and can be represented by the block-diagram depicted in Fig. \ref{TITOP}b. The TITOP model is composed of the direct dynamic model (transfer from acceleration twist to wrench) at the point $P_i$ and the inverse dynamic model (transfer from wrench to acceleration twist) at the point $C_i$.

As described in \cite{CHEBBI}, the state-space representation of $\mathfrak{T}_{P_i C_i}^{\mathcal{A}_i}(\mathrm{s})$ can be directly built from the following generalized second order model:
\begin{equation}
\begin{aligned}
& \underbrace{\left[\def\arraystretch{1.3}\begin{array}{cc}
\mathbf{D}^{\mathcal{A}_i}_{P_i} & {\mathbf{L}^{\mathcal{A}_i}_{P_i}}^{\mathrm T} \\
\mathbf{L}^{\mathcal{A}_i}_{P_i} & \mathbf{I}_n
\end{array}\right]}_{\mathbf{M}}\left[\begin{array}{c}
\ddot{\mathbf{x}}_{P_i} \\
\ddot{\boldsymbol{\eta}}
\end{array}\right]+\underbrace{\left[\begin{array}{cc}
\mathbf{0} & \mathbf{0} \\
\mathbf{0} & \operatorname{diag}\left(2 \xi^{\mathcal{A}_i}_j \omega^{\mathcal{A}_i}_j\right)
\end{array}\right]}_{\mathbf{D}}\left[\begin{array}{c}
\dot{\mathbf{x}}_{P_i} \\
\dot{\boldsymbol{\eta}}
\end{array}\right]+ \underbrace{\left[\begin{array}{cc}
\mathbf{0} & \mathbf{0} \\
\mathbf{0} & \operatorname{diag}\left({\omega_j^{\mathcal{A}_i}}^2\right)
\end{array}\right]}_{\mathbf{K}}\left[\begin{array}{c}
\mathbf{x}_{P_i} \\
\boldsymbol{\eta}
\end{array}\right] \\
& =\underbrace{\left[\def\arraystretch{1.5}\begin{array}{cc}
-\mathbf{I}_6 & \boldsymbol{\tau}_{C_i P_i}^{\mathrm T} \\
\mathbf{0} & {\boldsymbol{\Phi}^{\mathcal{A}_i}_{C_i}}^{\mathrm T}
\end{array}\right]}_{\mathbf{B}}\left[\begin{array}{c}
\mathbf{W}_{\mathcal{A}_i/\mathcal{A}_{i-1},{P_{i}}} \\
\mathbf{W}_{\mathcal{A}_{i+1}/\mathcal{A}_{i},{C_{i}}}
\end{array}\right]
\label{titop_ss}
\end{aligned}
\end{equation}

where:
\begin{itemize}
\item $n$ is the number of flexible clamped-free modes of the body $\mathcal{A}_i$ characterized by the modal coordinates vector $\boldsymbol{\eta}$, the frequencies $\omega^{\mathcal{A}_i}_j$ and the damping ratios $\xi^{\mathcal{A}_i}_j$, for $j=1, \ldots, n$.
\item $\mathbf{L}^{\mathcal{A}_i}_{P_i}$ is the $n \times 6$ modal participation factor matrix of the body at point $P_i$.
\item $\boldsymbol{\Phi}^{\mathcal{A}_i}_{C_i}$ is the $6 \times n$ projection matrix of the $n$ clamped-free modal shapes on the 6 DOFs at point $C_i$.
\item $\mathbf{D}^{\mathcal{A}_i}_{P_i}=\left[\begin{array}{cc}m^{\mathcal{A}_i} \textbf{I}_{3} & \mathbf{0}_{3\times 3} \\ \mathbf{0}_{3\times 3} & \textbf{J}_{P_i}^{\mathcal{A}_i}\end{array}\right] $ is the $6 \times 6$ static direct dynamic model of the body at point $P_i$, where $m^{\mathcal{A}_i}$ is the mass of $\mathcal{A}_i$ and $\textbf{J}_{P_i}^{\mathcal{A}_i}$ represents the inertia tensor of $\mathcal{A}_i$ written in the body frame $\mathcal{R}_{a_i}$.
\end{itemize}

The right-hand term of Eq. \eqref{titop_ss} describes the contribution of the wrenches $\mathbf{W}_{\mathcal{A}_i/\mathcal{A}_{i-1},{P_{i}}}$ and $\mathbf{W}_{\mathcal{A}_{i+1}/\mathcal{A}_{i},{C_{i}}}$ to the generalized force vector. The acceleration twist $\ddot{\mathbf{x}}_{C_i}$ at point $C_i$ can be easily expressed from the acceleration twist $\ddot{\mathbf{x}}_{P_i}$ at point $P_i$, the generalized acceleration vector $\ddot{\boldsymbol{\eta}}$, the modal shapes $\boldsymbol{\Phi}^{\mathcal{A}_i}_{C_i}$ and the kinematic model $\boldsymbol{\tau}_{C_i P_i}$, as follows:

\begin{equation}
\ddot{\mathbf{x}}_{C_i}=\boldsymbol{\Phi}^{\mathcal{A}_i}_{C_i} {\ddot{\boldsymbol{\eta}}}+\boldsymbol{\tau}_{C_i P_i} \ddot{\mathbf{x}}_{P_i}
\label{xCi}
\end{equation}

Finally, from the first row of Eq. \eqref{titop_ss}, the output equation for the wrench $\mathbf{W}_{\mathcal{A}_i/\mathcal{A}_{i-1},{P_{i}}}$ is equal to:

\begin{equation}
\mathbf{W}_{\mathcal{A}_i/\mathcal{A}_{i-1},{P_{i}}}=-{\mathbf{L}^{\mathcal{A}_i}_{P_i}}^{\mathrm T} {\ddot{\boldsymbol{\eta}}}-\mathbf{D}^{\mathcal{A}_i}_{P_i} \ddot{\mathbf{x}}_{P_i}+\boldsymbol{\tau}_{C_i P_i}^{\mathrm T} \mathbf{W}_{\mathcal{A}_{i+1}/\mathcal{A}_{i},{C_{i}}}
\label{WAP}
\end{equation}

From Eqs. \eqref{titop_ss}, \eqref{xCi} and \eqref{WAP}, the general state-space realization of the TITOP model $\mathfrak{T}_{P_i C_i}^{\mathcal{A}_i}(\mathrm{s})$ \cite{finozzi} reads:
\begin{equation}
\left[\begin{array}{c}
\dot{\boldsymbol{\eta}} \\
\ddot{\boldsymbol{\eta}} \\
\hline \ddot{\mathbf{x}}_{C_i} \\
\mathbf{W}_{\mathcal{A}_i/\mathcal{A}_{i-1},{P_{i}}}
\end{array}\right]=\left[\begin{array}{c|c}\mathbf{A}&\mathbf{B}\\\hline\mathbf{C}&\mathbf{D}\end{array}\right]\left[\begin{array}{c}
\boldsymbol{\eta} \\
\dot{\boldsymbol{\eta}} \\
\hline \mathbf{W}_{\mathcal{A}_{i+1}/\mathcal{A}_{i},{C_{i}}} \\
\ddot{\mathbf{x}}_{P_i}
\end{array}\right] 
\end{equation}

where the matrices $\mathbf{A}$, $\mathbf{B}$, $\mathbf{C}$ and $\mathbf{D}$ are given by:

\begin{equation}
\begin{aligned}
&\mathbf{A}=\left[\begin{array}{cc}\mathbf{0}_{n\times n}&\mathbf{I}_{n}\\-\operatorname{diag}\left({\omega^{\mathcal{A}_i}_j}^2\right) & -\operatorname{diag}\left(2 \xi^{\mathcal{A}_i}_j \omega^{\mathcal{A}_i}_j\right)\end{array}\right]_{\mathcal{R}_{a_i}}\\
&\mathbf{B}=\left[\begin{array}{cc}\mathbf{0}_{n\times6}&\mathbf{0}_{n\times6}\\{\boldsymbol{\Phi}^{\mathcal{A}_i}_{C_i}}^{\mathrm T} & -\mathbf{L}^{\mathcal{A}_i}_{P_i}\end{array}\right]_{\mathcal{R}_{a_i}}\\
& \mathbf{C}=\left[\begin{array}{cc}-\boldsymbol{\Phi}^{\mathcal{A}_i}_{C_i} \operatorname{diag}\left({\omega^{\mathcal{A}_i}_j}^2\right) & -\boldsymbol{\Phi}^{\mathcal{A}_i}_{C_i} \operatorname{diag}\left(2 \xi^{\mathcal{A}_i}_j \omega^{\mathcal{A}_i}_j\right)\\{\mathbf{L}^{\mathcal{A}_i}_{P_i}}^{\mathrm T} \operatorname{diag}\left({\omega^{\mathcal{A}_i}_j}^2\right) & {\mathbf{L}^{\mathcal{A}_i}_{P_i}}^{\mathrm T} \operatorname{diag}\left(2 \xi^{\mathcal{A}_i}_j \omega^{\mathcal{A}_i}_j\right)\end{array}\right]_{\mathcal{R}_{a_i}}\\
&\mathbf{D}=\left[\begin{array}{cc}\boldsymbol{\Phi}^{\mathcal{A}_i}_{C_i} {\boldsymbol{\Phi}^{\mathcal{A}_i}_{C_i}}^{\mathrm T} & \left({\boldsymbol{\tau}}_{C_i P_i}-\boldsymbol{\Phi}^{\mathcal{A}_i}_{C_i} \mathbf{L}^{\mathcal{A}_i}_{P_i}\right)\\\left({\boldsymbol{\tau}}_{C_i P_i}-\boldsymbol{\Phi}^{\mathcal{A}_i}_{C_i} \mathbf{L}^{\mathcal{A}_i}_{P_i}\right)^{\mathrm T} & \underbrace{-\mathbf{D}^{\mathcal{A}_i}_{P_i}+{\mathbf{L}^{\mathcal{A}_i}_{P_i}} ^{\mathrm T}\mathbf{L}^{\mathcal{A}_i}_{P_i}}_{-\mathbf{R}^{\mathcal{A}_i}_{P_i}}\end{array}\right]_{\mathcal{R}_{a_i}}\end{aligned}
\end{equation}

One can also introduce the residual mass matrix $\mathbf{R}^{\mathcal{A}_i}_{P_i}$ of the body $\mathcal{A}_i$ at point $P_i$, which is always definite positive by definition:
\begin{equation}
\mathbf{R}^{\mathcal{A}_i}_{P_i}=\mathbf{D}^{\mathcal{A}_i}_{P_i}-{\mathbf{L}^{\mathcal{A}_i}_{P_i}}^{\mathrm T} \mathbf{L}^{\mathcal{A}_i}_{P_i}
\end{equation}

If only the direct dynamic model at the point $P_i$ is considered, the TITOP model is denoted as  ${\left[\mathfrak{T}_{P_{i} }^{\mathcal{A}_{i}}(\mathrm{s})\right]}_{\mathcal{R}_{a_i}}$.

\subsection{Analytical $n$-port model of a rigid body}
Let us consider a general rigid body $\mathcal{B}$ with center of mass $G$. Considering that the rigid body is submitted to external forces/moments $\mathbf{F}_{\mathrm{ext} / \mathcal{B}, G}, \mathbf{T}_{\mathrm{ext} / \mathcal{B}, G}$ (i.e. solar radiation pressure, gravity gradient, atmospheric drag, magnetic fields, etc.) and to forces/moments
$\mathbf{F}_{\mathcal{B} / \mathcal{A}, P}, \mathbf{T}_{\mathcal{B}/ \mathcal{A}, P}$ due to interactions with an appendage $\mathcal{A}$ connected at point $P$, the linearized Newton-Euler equations read:

\begin{equation}
\underbrace{\left[\begin{array}{c}\mathbf{F}_{\mathrm{ext} / \mathcal{B}, G}-\mathbf{F}_{\mathcal{B} / \mathcal{A},G} \\ \mathbf{T}_{\mathrm{ext} / \mathcal{B}, G}-\mathbf{T}_{\mathcal{B} / \mathcal{A},G}\end{array}\right]}_{\mathbf{W}_{\mathrm{ext} / \mathcal{B}, G}-\mathbf{W}_{\mathcal{B} / \mathcal{A}, G}}=\textbf{D}_{G}^{\mathcal{B}}\underbrace{\left[\begin{array}{c}\mathbf{a}_{G} \\ \boldsymbol{\dot{\omega}}_{G}\end{array}\right]}_{\ddot{\mathbf{x}}_{G}}, \text { with } \quad \textbf{D}_{G}^{\mathcal{B}}=\left[\begin{array}{cc}m^{\mathcal{B}} \textbf{I}_{3} & \mathbf{0}_{3\times 3} \\ \mathbf{0}_{3\times 3} & \textbf{J}_{G}^{\mathcal{B}}\end{array}\right] 
\label{rheq}
\end{equation}

The mathematical relation between the wrenches $\mathbf{W}_{\mathcal{B} / \mathcal{A}, G}$, $\mathbf{W}_{\mathcal{B} / \mathcal{A}, P}$ and the acceleration twists ${\ddot{\mathbf{x}}_{G}}$, ${\ddot{\mathbf{x}}_{P}}$ is given by:

\begin{equation}
\underbrace{\left[\begin{array}{c} \mathbf{F}_{\mathcal{B} / \mathcal{A}, G} \\ \mathbf{T}_{\mathcal{B} / \mathcal{A}, G} \end{array}\right]}_{\mathbf{W}_{\mathcal{B} / \mathcal{A}, G}}={\boldsymbol{\tau}^{\mathrm{T}}_{PG}}\underbrace{\left[\begin{array}{c}
\mathbf{F}_{\mathcal{B} / \mathcal{A}, P} \\
\mathbf{T}_{\mathcal{B} / \mathcal{A}, P}
\end{array}\right]}_{\mathbf{W}_{\mathcal{B} / \mathcal{A}, P}} \quad \text { and } \quad \underbrace{\left[\begin{array}{c}\mathbf{a}_P \\ \boldsymbol{\dot{\omega}}_P\end{array}\right]}_{\ddot{\mathbf{x}}_{P}}={{\boldsymbol{\tau}}_{P G}} \underbrace{\left[\begin{array}{c}\mathbf{a}_G \\ \boldsymbol{\dot{\omega}}_G\end{array}\right]}_{\ddot{\mathbf{x}}_{G}}
\label{eqwrenches}
\end{equation}

From Eqs. \eqref{rheq} and \eqref{eqwrenches}, the model of a rigid body $\mathcal{B}$ connected to $n$ appendages can directly be obtained:

\begin{equation}
\left[\begin{array}{c}\ddot{\mathbf{x}}_{P_1} \\ \ddot{\mathbf{x}}_{P_2} \\ \vdots \\ \ddot{\mathbf{x}}_{P_n} \\ \ddot{\mathbf{x}}_G\end{array}\right]=\underbrace{\left[\begin{array}{c}{\boldsymbol{\tau}}_{P_1 G} \\ {\boldsymbol{\tau}}_{P_2 G} \\ \vdots \\ {\boldsymbol{\tau}}_{P_n G} \\ \mathbf{I}_6\end{array}\right]\left[\textbf{D}_{G}^{\mathcal{B}}\right]^{-1}\left[\begin{array}{lllll}{\boldsymbol{\tau}}_{P_1 G}^{\mathrm{T}} & {\boldsymbol{\tau}}_{P_2 G}^{\mathrm{T}} & \cdots & {\boldsymbol{\tau}}_{P_n G}^{\mathrm{T}} & \mathbf{I}_6\end{array}\right]}_{{\mathfrak{X}}_{P_1\ldots P_nG}^{\mathcal{B}}}\left[\begin{array}{c}\mathbf{W}_{\mathcal{A}_1 / \mathcal{B}, P_1} \\ \mathbf{W}_{\mathcal{A}_2 / \mathcal{B}, P_2} \\ \vdots \\ \mathbf{W}_{\mathcal{A}_{n} / \mathcal{B}, P_n} \\ \mathbf{W}_{\mathrm{ext} / \mathcal{B}, G}\end{array}\right]
\label{finalmodelrigidbody}
\end{equation}

where ${{\mathfrak{X}}_{P_1\ldots P_nG}^{\mathcal{B}}}$ is the $n$-port inverse linearized dynamic model of the body $\mathcal{B}$ computed at the points ${P_1, \ldots, P_n, G}$ in the body frame $\mathcal{R}_{b}$. This model can also be computed when one of the ports is inverted (transfer from acceleration twist to wrench). For instance, if the first port is inverted, the model is denoted as:

\begin{equation}
\left[\begin{array}{c}\mathbf{W}_{\mathcal{B} / \mathcal{A}_1, P_1} \\ \ddot{\mathbf{x}}_{P_2} \\ \vdots \\ \ddot{\mathbf{x}}_{P_n} \\ \ddot{\mathbf{x}}_G\end{array}\right]=\underbrace{\left[\begin{array}{c}{\boldsymbol{\tau}}_{G P_1}^{\mathrm{T}}\left[\begin{array}{lllll}{-\textbf{D}_{G}^{\mathcal{B}} {\boldsymbol{\tau}}_{G P_1}} & {\boldsymbol{\tau}}_{P_2 G}^{\mathrm{T}} & \cdots & {\boldsymbol{\tau}}_{P_n G}^{\mathrm{T}} & \mathbf{I}_6\end{array}\right] \\ \left[\begin{array}{c}{\boldsymbol{\tau}}_{P_2 G} \\ \vdots \\ {\boldsymbol{\tau}}_{P_n G} \\ \mathbf{I}_{6}\end{array}\right]\left[\begin{array}{lllll}{\boldsymbol{\tau}}_{G P_1} & \mathbf{0}_{6\times 6} & \cdots & \mathbf{0}_{6\times 6} & \mathbf{0}_{6\times 6}\end{array}\right]\end{array}\right]}_{{\mathfrak{R}}_{P_1\ldots P_nG}^{\mathcal{B}}}\left[\begin{array}{c}\ddot{\mathbf{x}}_{P_1} \\ \mathbf{W}_{\mathcal{A}_2 / \mathcal{B}, P_2} \\ \vdots \\ \mathbf{W}_{\mathcal{A}_{n} / \mathcal{B}, P_n} \\ \mathbf{W}_{\mathrm{ext} / \mathcal{B}, G}\end{array}\right]
\label{finalmodelrigidbody3}
\end{equation}

where ${{\mathfrak{R}}_{P_1\ldots P_nG}^{\mathcal{B}}}$ represents the $n$-port direct/inverse linearized dynamic model of the rigid body $\mathcal{B}$: the inverse linearized dynamic model computed at the points ${P_2, \ldots, P_n, G}$ and the linearized direct dynamic model (with a minus sign) computed at point $P_1$ in the body frame $\mathcal{R}_{b}$.

\subsection{Connection model between two different bodies}

In order to assemble linear TITOP/$n$-port models between each other, the Direct Cosine Matrix (DCM) $\mathbf{P}_{\mathcal{R}_a / \mathcal{R}_b}$ between the frame $\mathcal{R}_a$ attached to the body $\mathcal{A}$ and the frame $\mathcal{R}_b$ attached to the body $\mathcal{B}$ must be taken into account in the propagation of the wrenches and acceleration twists. Let $P$ be the point where $\mathcal{A}$ is connected to $\mathcal{B}$. If the body $\mathcal{A}$ is fixed in a certain angular orientation with respect to the body $\mathcal{B}$, it follows that:

\begin{equation}
\left[ \mathbf{W}_{\mathcal{A} / \mathcal{B}, P}\right]_{\mathcal{R}_b}=\underbrace{\operatorname{diag}\left(\mathbf{P}_{\mathcal{R}_a / \mathcal{R}_b}, \mathbf{P}_{\mathcal{R}_a / \mathcal{R}_b}\right)}_{\mathbf{P}_{\mathcal{R}_a / \mathcal{R}_b}^{\times 2}}\left[ \mathbf{W}_{\mathcal{A} / \mathcal{B}, P}\right]_{\mathcal{R}_a} \quad \text{and} \quad \left[ \ddot{\mathbf{x}}_{P}\right]_{\mathcal{R}_b}=\underbrace{\operatorname{diag}\left(\mathbf{P}_{\mathcal{R}_a / \mathcal{R}_b}, \mathbf{P}_{\mathcal{R}_a / \mathcal{R}_b}\right)}_{\mathbf{P}_{\mathcal{R}_a / \mathcal{R}_b}^{\times 2}}\left[ \ddot{\mathbf{x}}_{P}\right]_{\mathcal{R}_a}
\end{equation}

where the DCM $\mathbf{P}_{\mathcal{R}_a / \mathcal{R}_b}$ has a constant value. However, if the body $\mathcal{A}$ rotates around a certain direction vector $\mathbf{v}$, the required transformation matrix has to be parameterized according to the time-varying rotation angle $\alpha$. In this case, the rotation matrix will be represented as $\mathbf{P}(\alpha)$. As an example, if the unit vectors $\mathbf{z}_a$ and $\mathbf{z}_b$ are aligned and the appendage $\mathcal{A}$ rotates around $\mathbf{v} \equiv \mathbf{z}_a \equiv \mathbf{z}_b$, $\mathbf{P}(\alpha)$ can be written as follows:

\begin{equation}
\mathbf{P}(\alpha)=\left[\begin{array}{ccc}
\cos(\alpha) & -\sin(\alpha) & 0 \\
\sin(\alpha) & \cos(\alpha) & 0 \\
0 & 0 & 1
\end{array}\right] 
\label{DCM}
\end{equation}

\begin{figure}[!ht]
\centering
 \includegraphics[width=1\textwidth]{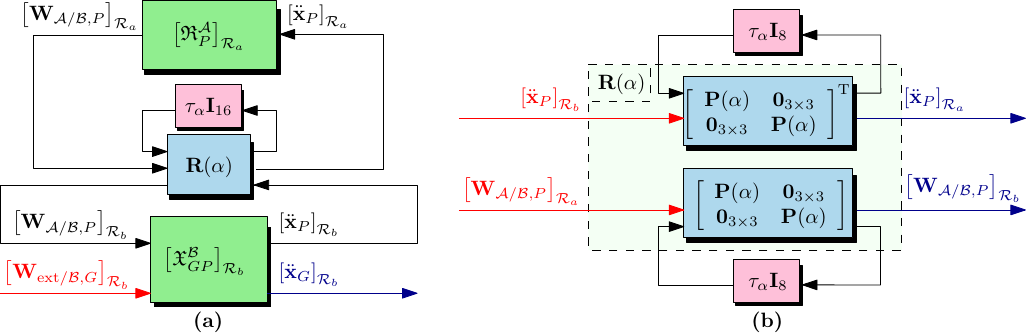}
\caption{(a) Block-diagram model of the connection between two rigid bodies $\mathcal{B}$ and $\mathcal{A}$. (b) Assembly of the transformation model $\mathbf{R}({\alpha})$.}
\label{connection} 
\end{figure}

An LFT parameterization of the rotation matrix described in Eq. \eqref{DCM} is implemented as demonstrated by Guy \textit{et al.} \cite{Guy2014}, with $\tau_{\alpha}=\tan (\alpha / 4)$ and $\tau_{\alpha} \in [-1, 1]$. The complete block-diagram model of the connection between two rigid bodies $\mathcal{B}$ and $\mathcal{A}$ is displayed in Fig. \ref{connection}a. The transformation model $\mathbf{R}({\alpha})$ can be observed in Fig. \ref{connection}b. Similarly, the constant transformation model $\mathbf{R}_{\mathcal{R}_a / \mathcal{R}_b}$ can also be derived from $\mathbf{P}_{\mathcal{R}_a / \mathcal{R}_b}$.

\subsection{Robotic arm model}

\begin{figure}[!ht]
\centering
 \includegraphics[width=1\textwidth]{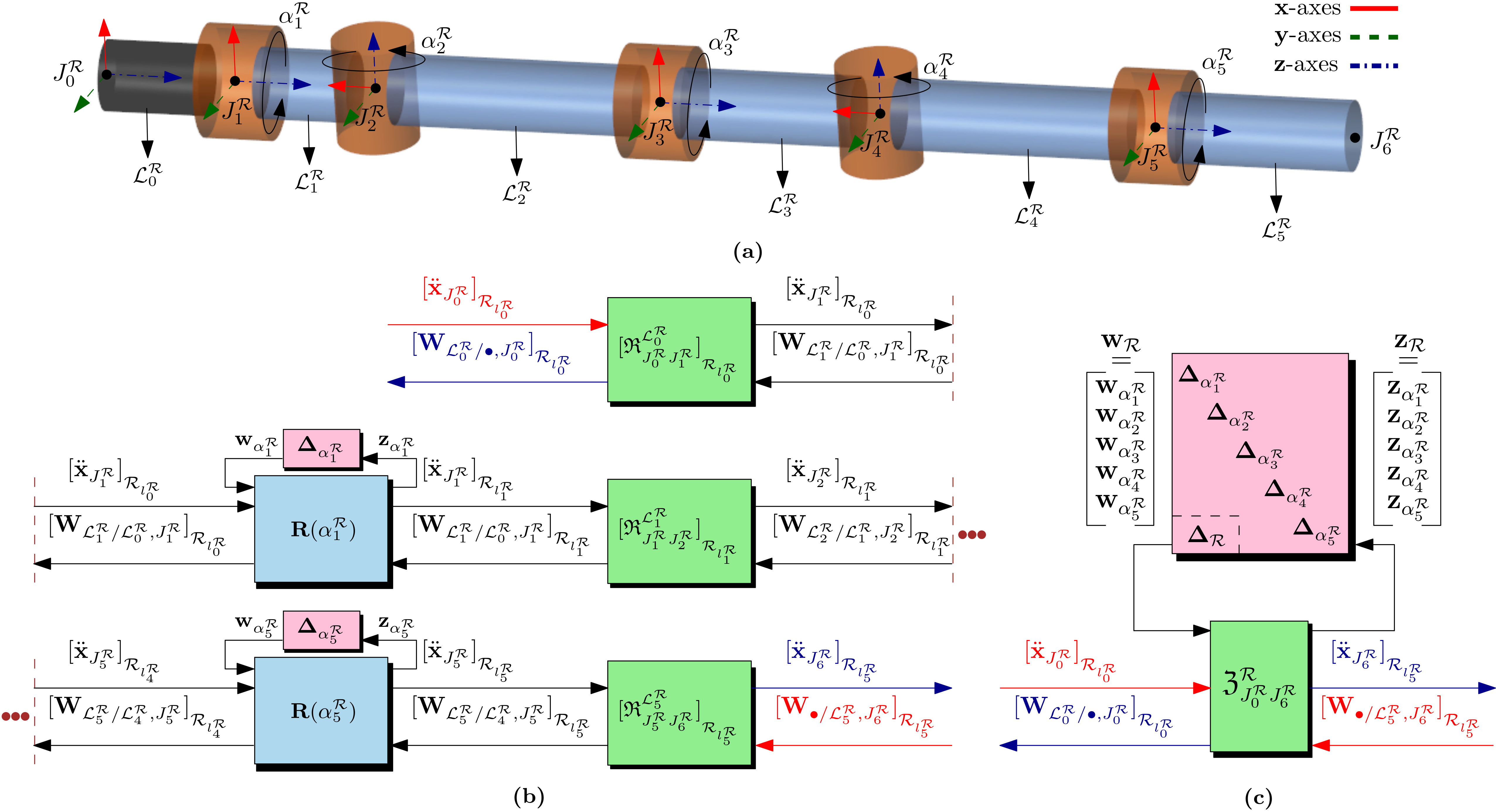}
\caption{Robotic manipulator representation: (a) robotic arm kinematics (Note: for the sake of simplicity, the $\mathbf{x}$-axes are displayed in solid red lines, the $\mathbf{y}$-axes in dashed green lines and the $\mathbf{z}$-axes in dash-dotted blue lines). (b) block-diagram of the parameterized robotic arm written in LFR form. (c) equivalent LFR form of the manipulator.}
\label{roboticarm} 
\end{figure}

Ultimately, the spacecraft uses a multi-arm robot with three similar robotic arms to assemble a flexible structure on-orbit. Let us now analyze the dynamics of one robotic arm, which is going to be denoted as $\mathcal{R}$. First, it should be noted that $\alpha_{k}^{\mathcal{R}}$ (for $k=1, \ldots,5$) represents the angular configuration of $\mathcal{R}$, as depicted in Fig. \ref{roboticarm}a. In addition, the robotic arm's 6 different links $\mathcal{L}^{\mathcal{R}}_{i}$ (for $i=0, \ldots,5$) are assumed to be rigid. The TITOP dynamic model of each link $\mathcal{L}^{\mathcal{R}}_{i}$ is given by ${[\mathfrak{R}_{J^{\mathcal{R}}_{i}J^{\mathcal{R}}_{i+1}}^{\mathcal{L}^{\mathcal{R}}_i}]}_{\mathcal{R}_{l^{\mathcal{R}}_i}}$, with $\mathcal{R}_{l^{\mathcal{R}}_{i}}=\left({J^{\mathcal{R}}_{i}}; \mathbf{x}_{l^{\mathcal{R}}_{i}}, \mathbf{y}_{l^{\mathcal{R}}_{i}}, \mathbf{z}_{l^{\mathcal{R}}_{i}}\right)$, since  $\mathcal{L}^{\mathcal{R}}_{i}$ is connected to a parent substructure at point $J^{\mathcal{R}}_{i}$ and to a child substructure at point $J^{\mathcal{R}}_{i+1}$. 

\textbf{Varying rotation angles of the robotic arm} $\alpha^{\mathcal{R}}_{k}$: The robotic arm model is parameterized according to the manipulator's geometrical configuration $\alpha^{\mathcal{R}}_{k}$, respectively defined in the reference frames $\mathcal{R}_{l^{\mathcal{R}}_{k}}$, as displayed in Fig. \ref{roboticarm}a. The uncertainty block describing the changing geometrical configuration of the robotic arm $\mathcal{R}$ is given by $\boldsymbol{\Delta}_{\mathcal{R}}=\operatorname{diag}\left(\boldsymbol{\Delta}_{\alpha^{\mathcal{R}}_{1}}, \boldsymbol{\Delta}_{\alpha^{\mathcal{R}}_{2}}, \boldsymbol{\Delta}_{\alpha^{\mathcal{R}}_{3}}, \boldsymbol{\Delta}_{\alpha^{\mathcal{R}}_{4}},
\boldsymbol{\Delta}_{\alpha^{\mathcal{R}}_{5}}\right)$, with $\boldsymbol{\Delta}_{\alpha^{\mathcal{R}}_{k}}=\tau_{\alpha^{\mathcal{R}}_{k}}I_{16}$ and $\tau_{\alpha^{\mathcal{R}}_{k}} \in [-1, 1]$. Fig. \ref{roboticarm}b depicts the block-diagram of the parameterized robotic arm written in LFR form, whereas Fig. \ref{roboticarm}c displays the equivalent global LFR form of the robotic arm, where $\mathfrak{Z}^{\mathcal{R}}_{J^{\mathcal{R}}_0 J^{\mathcal{R}}_6}$ is the dynamic model of $\mathcal{R}$ and $\mathbf{w}_{\mathcal{R}}\boldsymbol{\Delta}_{\mathcal{R}}=\mathbf{z}_{\mathcal{R}}$. Furthermore, $\mathbf{w}_{\mathcal{R}}$ and $\mathbf{z}_{\mathcal{R}}$ are the endogenous inputs and outputs of the arm manipulator model. It should also be noted that the angular configuration of the arm represented in Fig. \ref{roboticarm}a is given by $\alpha^{\mathcal{R}}_{k}=0$ \si{\radian}.

The dynamic model $\left[\mathfrak{Z}^{\mathcal{R}}_{J^{\mathcal{R}}_0 J^{\mathcal{R}}_6}\right]^{-1}$ can also be obtained by inverting the channels from ${[\ddot{\mathbf{x}}_{J^{\mathcal{R}}_0}]}_{\mathcal{R}_{l^{\mathcal{R}}_0}}$ to ${[\mathbf{W}_{\mathcal{L}^{\mathcal{R}}_0/\bullet,J^{\mathcal{R}}_0}]}_{\mathcal{R}_{l^{\mathcal{R}}_0}}$ and from ${[\mathbf{W}_{\bullet/\mathcal{L}^{\mathcal{R}}_5,J^{\mathcal{R}}_6}]}_{\mathcal{R}_{l^{\mathcal{R}}_5}}$ to ${[\ddot{\mathbf{x}}_{J^{\mathcal{R}}_6}]}_{\mathcal{R}_{l^{\mathcal{R}}_5}}$.

\subsection{Complete model of the system}

\begin{figure}[!h]
\centering
 \includegraphics[width=1\textwidth]{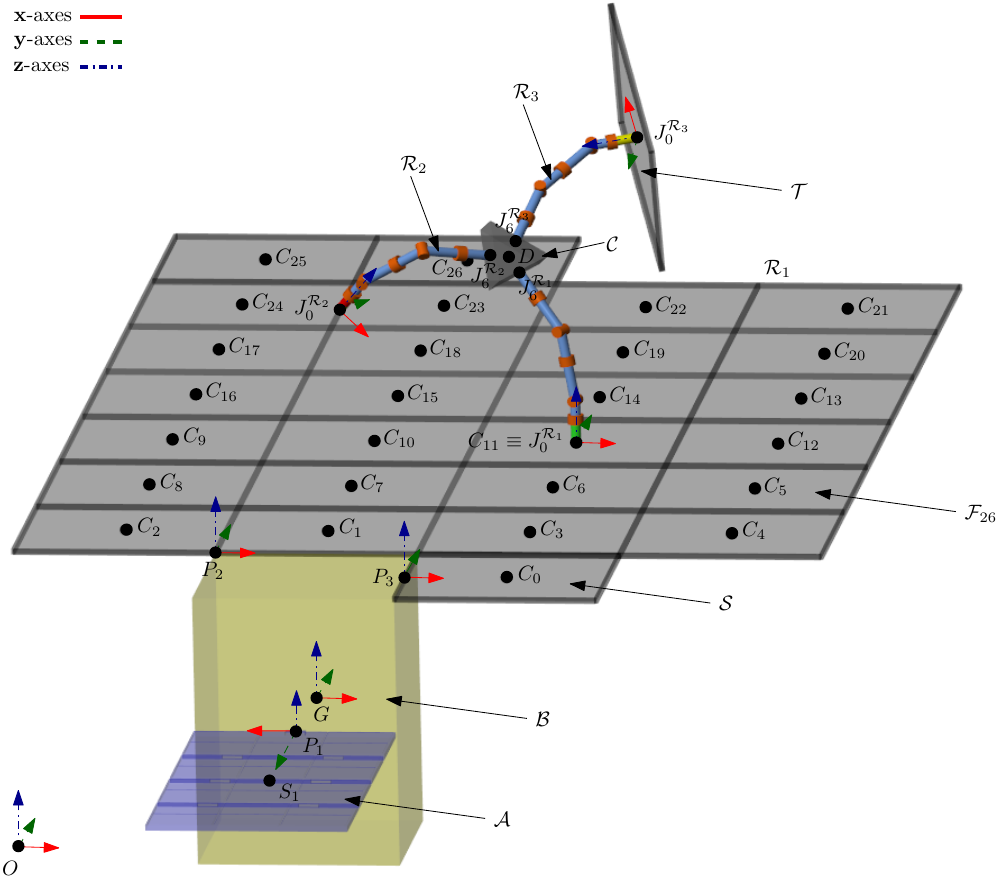}
\caption{3D representation of an on-orbit assembly mission scenario (Note: for the sake of simplicity, the
$\mathbf{x}$-axes are displayed in solid red lines, the $\mathbf{y}$-axes in dashed green lines and the $\mathbf{z}$-axes in dash-dotted blue lines).}
\label{COMPLETEMODEL_mirror_fig} 
\end{figure}

An on-orbit assembly mission scenario is considered, as depicted in Fig. \ref{COMPLETEMODEL_mirror_fig}, where $\mathcal{R}_{i}=\left(O; \mathbf{x}_{i}, \mathbf{y}_{i}, \mathbf{z}_{i}\right)$ represents the inertial frame. The spacecraft in consideration consists of a rigid hub $\mathcal{B}$, a flexible solar array $\mathcal{A}$ and a stack of modular tiles $\mathcal{S}$. By means of a multi-arm robot equipped with three robotic arms $\mathcal{R}_{1-3}$ and a central hub $\mathcal{C}$, a large flexible structure is assembled. During the on-orbit assembly, various structures can be formed, with the number of possible configurations equal to the number of tiles in the fully assembled structure $N$, assuming the assembly order of the tiles is predefined. As a result, the time-varying flexible structure is denoted as $\mathcal{F}_n$, for $n=1, \ldots,N$.

\begin{figure}[!h]
\centering
 \includegraphics[width=1\textwidth]{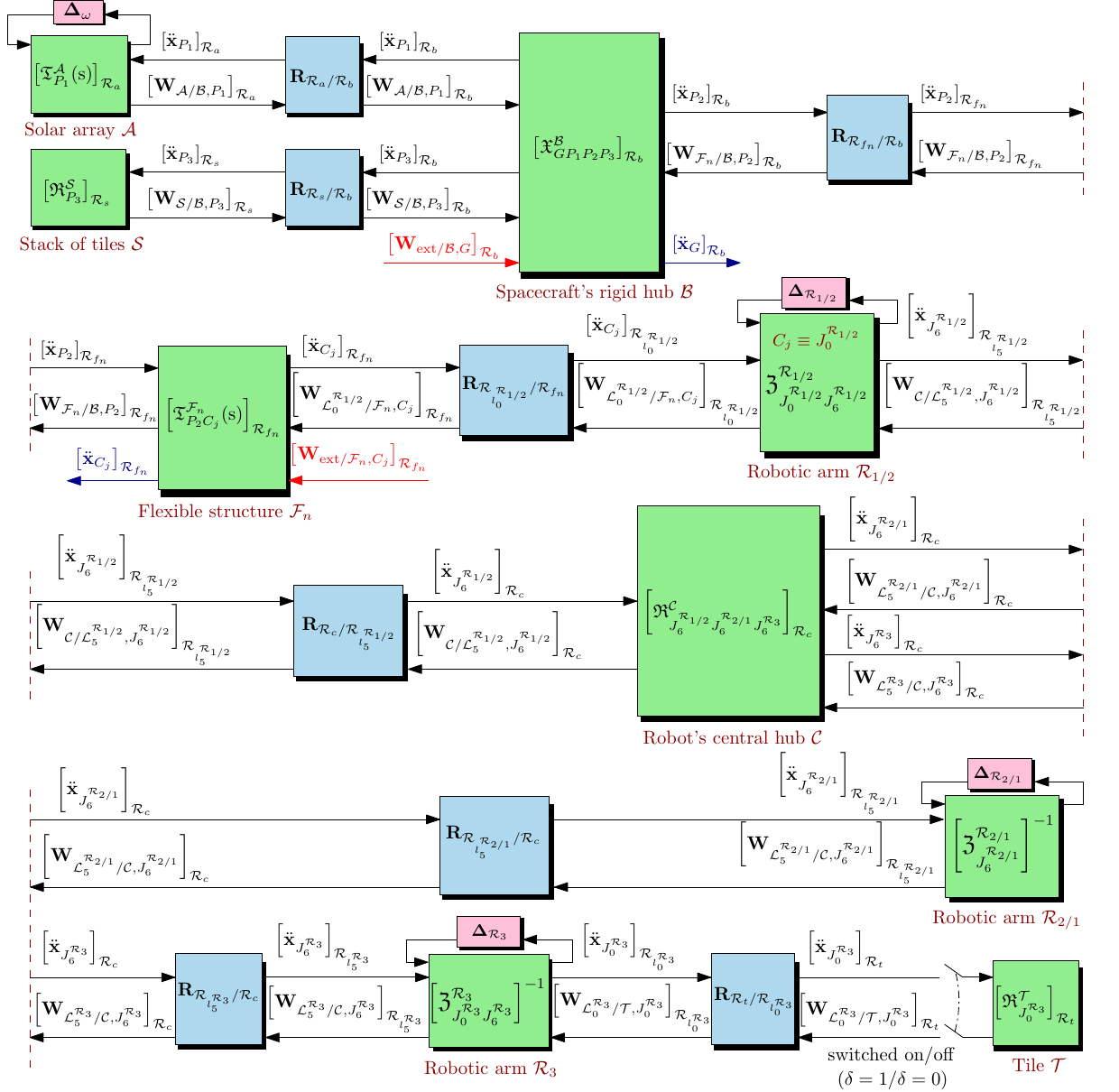}
\caption{Detailed block-diagram representation of an on-orbit assembly mission scenario.}
\label{COMPLETEMODEL_mirror_block} 
\end{figure}

The block-diagram depicted in Fig. \ref{COMPLETEMODEL_mirror_block} corresponds to the mission scenario shown in Fig. \ref{COMPLETEMODEL_mirror_fig}. In this block-diagram, the point $C_j$ associated with the TITOP model ${\left[\mathfrak{T}_{P_2 C_j}^{\mathcal{F}_n}(\mathrm{s}) \right]_{\mathcal{R}_{f_n}}}$ consistently represents the docking port where the multi-arm robot connects to the flexible structure. The options for this docking port logically depend on the state of the flexible structure being assembled, where the possible docking ports for the structure $\mathcal{F}_n$ are denoted by $C_j$, for $j=1, \ldots,n$. These docking ports are always located at the center of each modular tile, as shown in Fig. \ref{COMPLETEMODEL_mirror_fig}. It should also be noted that the port $C_j$, designed to facilitate robot traversal, is associated with the $\text{j}^{\text{th}}$ tile to be assembled.

As illustrated in the block-diagram in Fig. \ref{COMPLETEMODEL_mirror_block}, both robotic arms $\mathcal{R}_1$ (green end effector) and $\mathcal{R}_2$ (red end effector) can be connected to the flexible structure at various docking ports, as these arms are utilized for robot traversal. On the other hand, the third robotic arm $\mathcal{R}_3$ (yellow end effector) is dedicated solely to picking up and assembling tiles. 

To model the two different cases regarding whether the multi-arm robot is carrying a tile or not, two possibilities are considered. The first case can be seen in representation {\Large \textcircled{\normalsize 1}} in Fig. \ref{Diapositive1}, where the robot is not carrying a tile. In this case, all the six DOFs channels connecting the tile model ${\left[\mathfrak{R}_{J^{\mathcal{R}_3}_0}^{\mathcal{T}}\right]_{\mathcal{R}_{t}}}$ and the third robotic arm model $\left[{\mathfrak{Z}}^{\mathcal{R}_{3}}_{J^{\mathcal{R}_{3}}_0 J^{\mathcal{R}_{3}}_6}\right]^{-1}
$ are completely switched off.
 Moreover, the second case is illustrated in representations {\Large \textcircled{\normalsize 2}} and {\Large \textcircled{\normalsize 3}}, where the robot is carrying a tile. In this case, the same six DOFs channels are completely connected. This connection is visually represented in Fig. \ref{COMPLETEMODEL_mirror_block}, where $\mathcal{T}$ represents the tile being carried by the robot.

Initially, all the modular tiles used to assemble the flexible structure are mounted on the spacecraft's rigid body $\mathcal{B}$, awaiting assembly. The multi-arm robot is responsible for picking up and assembling the tiles one by one, causing the number of tiles in the stack to vary over time. Therefore, the mass and inertial properties of the stack of tiles can be represented as:

\begin{equation}
m^{S}=m^{T}(N-n-\delta) \quad \text{and} \quad \textbf{J}_{C_0}^{\mathcal{S}}=\textbf{J}_{J^{\mathcal{R}_3}_0}^{\mathcal{T}}(N-n-\delta)
\label{massunc}
\end{equation}

In Eq. \eqref{massunc}, the mass and inertia (computed at the stack's center of mass $C_0$) of the stack of tiles are calculated based on the mass and inertia of a single tile, taking into account the number of tiles already assembled and whether the robot is currently carrying a tile. Therefore, the variable $\delta$ equals 0 if the robot is not carrying a tile and 1 if it is. It can also be observed in Fig. \ref{COMPLETEMODEL_mirror_fig} that the stack of tiles is clamped to the spacecraft at the point $P_3$, while the flexible structure $\mathcal{F}_n$ is clamped to the satellite at the point $P_2$. Additionally, it can also be seen in Fig. \ref{COMPLETEMODEL_mirror_block} that both the tile $\mathcal{T}$ and the stack of tiles $\mathcal{S}$ are considered to be rigid bodies,  since the flexibility of the individual tiles is only relevant at high frequencies.

For any structure $\mathcal{F}_n$ and docking port $C_j$, a linear model can be computed. This model is also defined based on whether the robot is carrying a tile and which robotic arm is connected to the structure. The number of tiles in the stack depends on the state of the structure and whether the multi-arm robot is carrying a tile, as previously explained. Ultimately, the resulting model is fully parameterized according to the geometrical configuration of the multi-arm robot, given by $\boldsymbol{\Delta}_{\mathcal{R}_1}$, $\boldsymbol{\Delta}_{\mathcal{R}_2}$ and $\boldsymbol{\Delta}_{\mathcal{R}_3}$, as shown in Fig. \ref{COMPLETEMODEL_mirror_block}. 

\textbf{Modal uncertainty $\boldsymbol{\Delta}_{\omega}$}: Changes in structural parameters can provoke variations in the natural frequency of some flexible modes. Since the system's bandwidth of interest can be highly impacted by these parameters, relative uncertainty is taken into consideration on the natural frequency of the first flexible mode of the solar array $\mathcal{A}$, which is given by $\omega_{1}^\mathcal{{A}}$. Therefore, it follows that:

\begin{equation}
\omega_{1}^\mathcal{{A}}=\omega^{0\mathcal{A}}_{1}(1+r_{\omega_{1}^\mathcal{{A}}}\delta_{\omega_{1}^\mathcal{{A}}}) \end{equation}

where $\omega^{0\mathcal{A}}_{1}$ represents the nominal natural frequency, $\delta_{\omega_{1}^\mathcal{{A}}} \in [-1, 1]$ is a normalized real uncertainty and the parameter $r_{\omega_{1}^\mathcal{{A}}}$ is used to set the maximum percent of variation for $\omega_{1}^\mathcal{{A}}$. Since the uncertainty $\delta_{\omega_{1}^\mathcal{{A}}}$ appears two times for the first flexible mode in a minimal LFR of a flexible appendage, as explained in \cite{Guy2014}, the modal uncertainty block linked to the solar array $\mathcal{{A}}$ is equal to $\boldsymbol{\Delta}_{\omega}=\delta_{\omega_{1}^\mathcal{{A}}}\textbf{I}_{2}$, as depicted in Fig. \ref{COMPLETEMODEL_mirror_block}.

Taking into account all possible structures, docking ports, the connection of robotic arms to the structure and whether the robot carries a tile or not, there are $ 4 \sum_{c=1}^{N} c=4(N(N+1)/2) = 2N(N+1)$ different dynamic systems that require computation, such that a complete understanding of the system dynamics is achieved for the on-orbit assembly of a flexible structure composed of $N$ tiles (assuming the assembly order of the tiles is predefined). These models are global LFR representations, fully capturing the dynamics and interactions between all subsystems of the scenario being studied. Additionally, these models account for all the parameterization effects in a very compact representation. Fig. \ref{COMPLETEMODEL_mirror_block} illustrates the internal structure of all the possible overall LFR models and the interconnections between the several subsystems. In this representation, all block parameterizations are isolated at the component level. However, it should be noted that this set of $2N(N+1)$ dynamical systems does not account for scenarios involving a closed-loop kinematic chain. For instance, this includes cases where the robotic arms $\mathcal{R}_1$ and $\mathcal{R}_2$ are simultaneously connected to the flexible structure $\mathcal{F}_n$. For this reason, the notation $\mathcal{R}_{1/2}$ (or $\mathcal{R}_{2/1}$) used in the block-diagram shown in Fig. \ref{COMPLETEMODEL_mirror_block} indicates that if the robotic arm $\mathcal{R}_1$ is connected to the flexible structure, then the arm $\mathcal{R}_2$ is free, and vice versa.

This family of $2N(N+1)$ models can be easily built using the SDTlib \cite{userguide}, which enables the assembly of the complete system shown in Fig. \ref{COMPLETEMODEL_mirror_block} within a \textsc{Simulink} environment. Utilizing the SDTlib offers the possibility of declaring the parameters in each subsystem as uncertain and facilitates the rapid evaluation of all these models in the frequency domain. Without the SDTlib, accomplishing the proposed tasks would be extremely cumbersome and time-consuming. Ultimately, all the numerical values of the numerous system parameters which are employed in this section are described in Tables \ref{tab:Sat_prop} and \ref{tab:Sat_prop2}.

Having accurately modeled the dynamics of a three-legged robot assembling a large flexible structure, it is now important to understand the impact of this walking robot on the dynamics of the spacecraft. Furthermore, this knowledge can be utilized to determine the optimal path for the robot when assembling the large flexible structure.

\section{Path optimization for the on-orbit assembly of a large flexible structure using a multi-arm robot}

\label{pathopt}

As previously mentioned, there are $2N(N+1)$ different linear dynamic systems that can be computed by means of the block-diagram depicted in Fig. \ref{COMPLETEMODEL_mirror_block}. Knowing the assembly order of the structure, the objective of this section is to create a function that defines all the possible movements for the robot once it is connected to a specific tile of the structure and aims to move to a final objective point, assuming the robot can move, pick up a tile and/or assemble a tile sideways or diagonally. Once a map of all possible trajectories for the robot is created, it can be used to compute the robot's paths when picking up and assembling tiles, while avoiding scenarios that might lead to poor system performance and/or instability. To accomplish this, node graph theory \cite{williamsonlists} was chosen to model all the possible combinations between different positions of the robot in various structures.

\subsection{Directed node graphs}

A directed graph, or digraph, is a graph where the edges have orientations \cite{williamsonlists}. In one restricted but commonly used sense of the term, a directed graph is an ordered pair comprising:

\begin{itemize}

\item  a set of vertices (also called nodes or points);
\item  a set of edges (also called directed edges, directed links, directed lines, arrows or arcs) which are ordered pairs of vertices (that is, an edge is associated with two distinct vertices).
\end{itemize}

The edges of a graph define a symmetric relation on the vertices, called the adjacency relation. A graph is fully defined by its adjacency matrix $\mathbf{A}$, where $\mathbf{A}_{(g,h)}$ specifies the number of connections from vertex $g$ to vertex $h$. In a simple graph, $\mathbf{A}_{(g,h)}$ is either 0, indicating no connection, or 1, indicating a connection. Additionally, it should be noted that $\mathbf{A}_{(g,g)}=0$ because an edge in a simple graph cannot start and end at the same vertex.

For each of the $N$ possible flexible structures during the on-orbit assembly scenario, two directed node graphs are constructed, one for picking up a tile and the other one for assembling a tile. Therefore, the total number of node graphs which are needed for the full on-orbit assembly is equal to $2N$, considering that the assembly process begins with one tile already preassembled (illustration {\Large \textcircled{\normalsize 1}} in Fig. \ref{Diapositive1}) and that the node graphs for the final structure $\mathcal{F}_N$ are computed, even though there are theoretically no more tiles to assemble. However, these node graphs might still be necessary to compute a specific trajectory for the multi-arm robot. 

\begin{figure}[!ht]
\centering
 \includegraphics[width=1\textwidth]{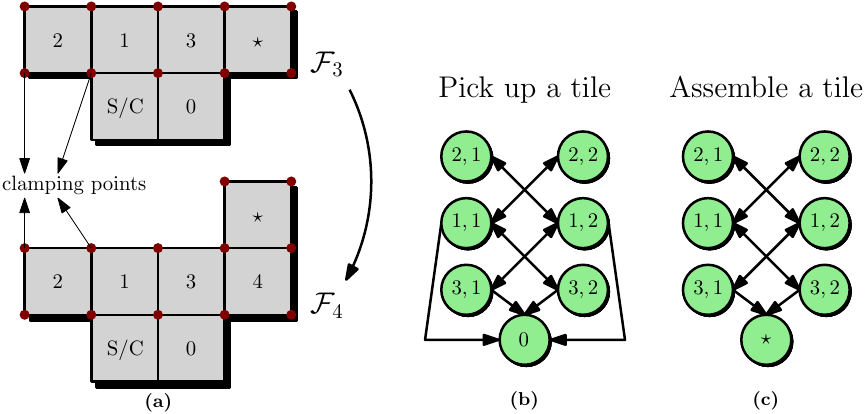}
\caption{Practical example of the on-orbit assembly scenario where the current flexible structure is composed of three tiles: (a) the current and subsequent states of the flexible structure as it is being assembled. (b) the node graph used to pick up a tile. (c) the node graph used to assemble a tile.}
\label{node_graph} 
\end{figure}

Figs. \ref{node_graph}b and \ref{node_graph}c illustrate an example of two node graphs used to compute the multi-arm robot's possible trajectories when the flexible structure being assembled comprises three different tiles (1, 2 and 3), as depicted in Fig. \ref{node_graph}a. In the graphs, the state of each node is represented by a pair of numbers, the first one indicating the tile number to which the multi-arm robot is connected and the second number representing which robotic arm is connected to the structure. In Figs. \ref{node_graph}a, \ref{node_graph}b and \ref{node_graph}c, the tile labeled with a 0 represents the stack from which the multi-arm robot can pick up tiles. Additionally, the tile labeled with a $\star$ indicates the position where the fourth tile should be placed by the robot and S/C represents the main hub $\mathcal{B}$ of the spacecraft. Logically, in this example, the main objective is for the multi-arm robot to assemble the fourth tile, thereby forming the structure $\mathcal{F}_4$. Therefore, the node graph shown in Fig. \ref{node_graph}b can initially be used to compute a trajectory for the robot to pick up a tile. Subsequently, the node graph in Fig. \ref{node_graph}c can be employed to compute the trajectory for the robot to assemble the tile it has picked up.

In the general case, the computation of the adjacency matrices for both node graphs assumes that the multi-arm robot can always move, pick up a tile and/or assemble a tile sideways or diagonally. In both node graphs, the adjacency matrices are $(2n+1)  \times (2n+1)$ square matrices, for $n=1, \ldots,N$, as there are $2n$ nodes related to the number of tiles in the structure $\mathcal{F}_n$ and another node that is either related to the stack of tiles or to the position of the next tile to be assembled.

For any structure $\mathcal{F}_n$, the data required for the model ${\mathfrak{T}_{P_2 C_j}^{\mathcal{F}_n}(\mathrm{s}) }$ can be obtained from NASTRAN by generating a finite element model (FEM) of the flexible structure, under the assumption that the body is clamped at the point $P_2$ and free at the point $C_j$. Whenever a tile is assembled to the structure, a new FEM is generated by clamping the new tile to the clamping points which are distributed along the existing structure, as depicted in Fig. \ref{node_graph}a for the example with three tiles. In Fig. \ref{node_graph}a, the docking ports $C_j$ (or central nodes) are omitted for simplicity, as they are represented in Fig. \ref{COMPLETEMODEL_mirror_fig}. Similarly, the clamping nodes are not shown in Fig. \ref{COMPLETEMODEL_mirror_fig} for the same reason.

Once all the necessary node graphs are built, weights can be added to all the edges connecting the different nodes. These weights can be computed based on various metrics, allowing for the definition of multiple different cost functions to generate these weights.

\subsection{Cost function engineering}

The main question addressed in this section is: how can different cost functions be designed in order to optimize stability, performance and control effort during the on-orbit assembly of flexible structures? These cost functions can be used to weight the several edges connecting the multiple nodes in the node graphs. The goal is to apply these cost functions to the linear dynamics associated with each edge connecting every pair of nodes. This is accomplished by operating in the frequency domain through the LPV formulation of the plant provided by the TITOP approach.

For example, in Fig. \ref{node_graph}, the node labeled \raisebox{.0pt}{\Large\textcircled{\raisebox{.9pt} {\scriptsize 1,1}}} represents the linear model composed of the flexible structure $\mathcal{F}_3$, where the robotic arm $\mathcal{R}_1$ is connected to the tile number 1. Depending on the node graph (Fig. \ref{node_graph}b or \ref{node_graph}c), the multi-arm robot might be holding a tile. However, in every case, this linear model is parameterized according to the geometrical configuration of the multi-arm robot.

\begin{figure}[!ht]
\centering
\includegraphics[width=1\textwidth]{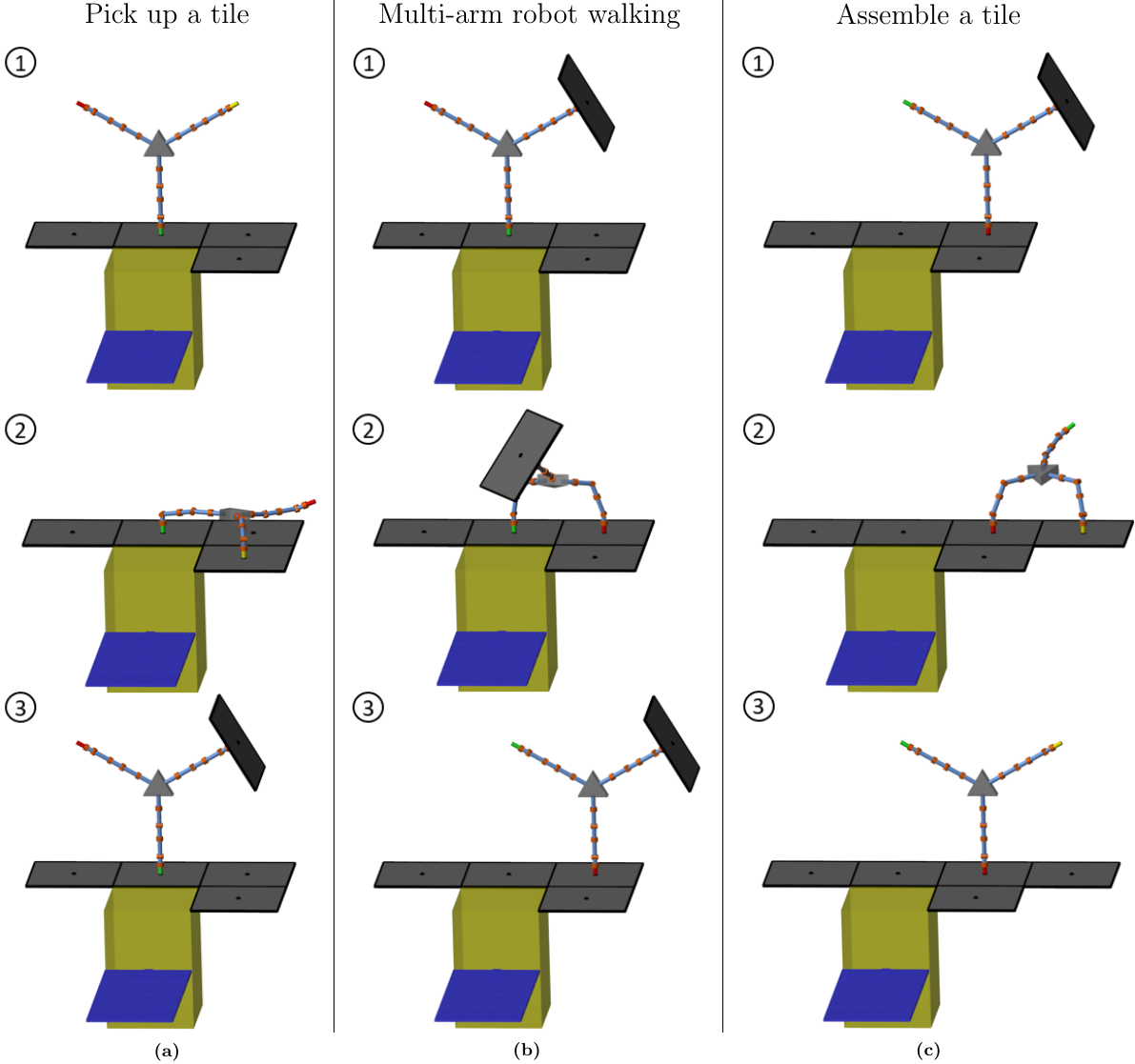}
\caption{Three different representations of the spacecraft for each of the three possible actions for the moving robot: (a) picking up a tile. (b) walking. (c) assembling a tile.}
\label{pickassemblemove} 
\end{figure}

Following the example shown in Fig. \ref{node_graph}, Fig. \ref{pickassemblemove} depicts three different representations of the three possible actions the multi-arm robot can perform: picking up a tile, walking and assembling a tile. Let us first look at the case where the multi-arm robot picks up a tile, as depicted in Fig. \ref{pickassemblemove}a. In order to weight the edge connecting the nodes given by the states \raisebox{.0pt}{\Large\textcircled{\raisebox{.9pt} {\scriptsize 1,1}}} and {\Large \textcircled{\normalsize 0}}, for example, two different trajectories are generated for the multi-arm robot joint angles, using fifth-order polynomials. The first trajectory is computed so that the initial position of the robot is given by $\alpha^{\mathcal{R}_{\bullet}}_{k}=0$ \si{\radian} (illustration {\Large \textcircled{\normalsize 1}} in Fig. \ref{pickassemblemove}a) and, at the end of this trajectory, the end effector of the robotic arm $\mathcal{R}_2$ aligns with the position of the stack of tiles (illustration {\Large \textcircled{\normalsize 2}} in Fig. \ref{pickassemblemove}a). Then, an array of $z$ models can be obtained by replacing $\boldsymbol{\Delta}_{\mathcal{R}_1}$, $\boldsymbol{\Delta}_{\mathcal{R}_2}$ and $\boldsymbol{\Delta}_{\mathcal{R}_3}$ (in the block-diagram depicted in Fig. \ref{COMPLETEMODEL_mirror_block}) according to a grid of $z$ equally distributed waypoints along this trajectory, resulting in an array of models $\mathbf{G}^m_1(\mathrm{s})$, for $m=1, \ldots,M$, where $M$ is the total number of possible robot paths/edges during the on-orbit assembly, including picking up a tile, walking and assembling a tile. In order to obtain $\mathbf{G}^m_1(\mathrm{s})$, the linear model corresponding to \raisebox{.0pt}{\Large\textcircled{\raisebox{.9pt} {\scriptsize 1,1}}} in the node graph depicted in Fig. \ref{node_graph}b is used, with $\delta=0$ (in the block-diagram depicted in Fig. \ref{COMPLETEMODEL_mirror_block}), since the robot is not carrying a tile. The second trajectory begins where the first one ends (illustration {\Large \textcircled{\normalsize 2}} in Fig. \ref{pickassemblemove}a) and concludes once again with $\alpha^{\mathcal{R}_{\bullet}}_{k}=0$ \si{\radian} (illustration {\Large \textcircled{\normalsize 3}} in Fig. \ref{pickassemblemove}a). In order to obtain the array of models $\mathbf{G}^m_2(\mathrm{s})$, which is obtained by replacing $\boldsymbol{\Delta}_{\mathcal{R}_1}$, $\boldsymbol{\Delta}_{\mathcal{R}_2}$ and $\boldsymbol{\Delta}_{\mathcal{R}_3}$ according to a grid of $z$ equally distributed waypoints along the second trajectory, the linear model corresponding to \raisebox{.0pt}{\Large\textcircled{\raisebox{.9pt} {\scriptsize 1,1}}} in the node graph depicted in Fig. \ref{node_graph}b is once again used, with $\delta=1$, since the robotic arm $\mathcal{R}_3$ is now carrying a tile. It is also important to mention that the initial geometrical configuration of the robot for the first trajectory and its final configuration for the second trajectory are consistently set to $\alpha^{\mathcal{R}_{\bullet}}_{k}=0$ \si{\radian} because this ensures uniform weighting of these edges, avoiding bias related to the initial and final positions of the multi-arm robot while performing an action.

A very similar procedure is followed when the multi-arm robot is walking (while carrying a tile or not), as can be observed in Fig. \ref{pickassemblemove}b. The only difference is that when replacing $\boldsymbol{\Delta}_{\mathcal{R}_1}$, $\boldsymbol{\Delta}_{\mathcal{R}_2}$ and $\boldsymbol{\Delta}_{\mathcal{R}_3}$ with the waypoints resulting from the computation of the second trajectory, the linear model corresponding to the node \raisebox{.0pt}{\Large\textcircled{\raisebox{.9pt} {\scriptsize 3,2}}} is used. In this scenario, it is also essential to ensure that the geometrical configuration of the robot depicted in illustration {\Large \textcircled{\normalsize 2}} in Fig. \ref{pickassemblemove}b remains unchanged when transitioning to the linear model associated with the node \raisebox{.0pt}{\Large\textcircled{\raisebox{.9pt} {\scriptsize 3,2}}}. Naturally, the linear models employed depend on the trajectories being analyzed, this is merely an example.

The same process applies when assembling a tile, as depicted in Fig. \ref{pickassemblemove}c. However, there is a small difference. For the linear model that is gridded over the second trajectory, the flexible structure $\mathcal{F}_4$ is considered, instead of $\mathcal{F}_3$, as a tile has been assembled. Logically, this change also deactivates the connection between the robotic arm $\mathcal{R}_3$ and $\mathcal{T}$, as the arm no longer holds a tile ($\delta=0$).

Ultimately, for each edge $m$, the arrays of models $\mathbf{G}^m_1(\mathrm{s})$ and $\mathbf{G}^m_2(\mathrm{s})$ can be concatenated into a single array of models containing $2z$ dynamical systems, denoted as $\mathbf{H}^m(\mathrm{s})$. For each edge, this array of dynamical systems $\mathbf{H}^m(\mathrm{s})$ precisely represents the open loop dynamics of the spacecraft when the multi-arm robot executes any path involving picking up a tile, walking or assembling a tile.

\begin{figure}[!ht]
\centering
\includegraphics[width=1\textwidth]{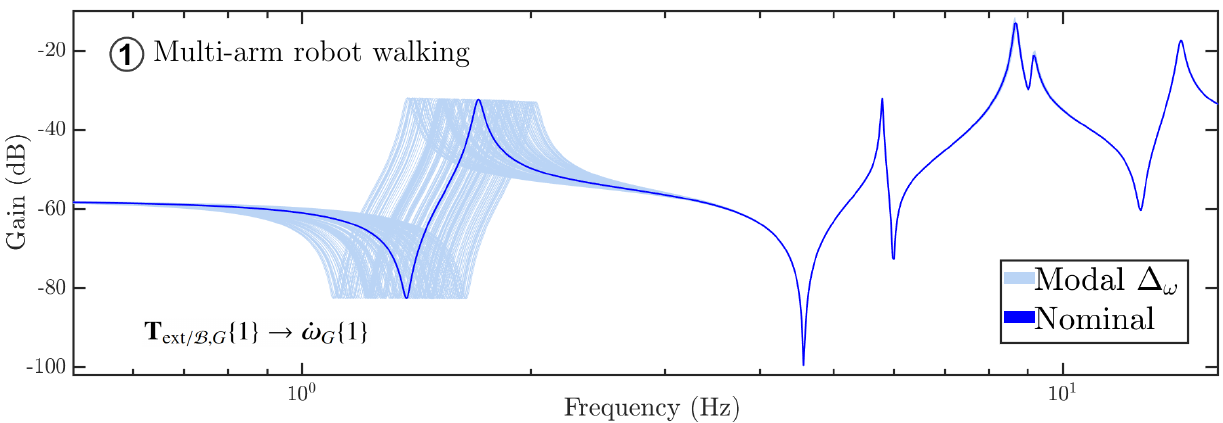}
\caption{Gains of the uncertain and nominal systems for the instant {\Large \textcircled{\normalsize 1}} displayed in Fig. \ref{pickassemblemove}b (${\mathbf{T}_{\mathrm{ext} / \mathcal{B}, G}}\{1\} \rightarrow \boldsymbol{\dot{{\omega}}}_{G}\{1\}$ transfer function).}
\label{singvalues} 
\end{figure}

By setting the varying parameters $\alpha^{\mathcal{R}_{\bullet}}_{k}$ and $\delta$ accordingly, the singular values of the system shown in Fig. \ref{COMPLETEMODEL_mirror_block} are analyzed at the instant {\Large \textcircled{\normalsize 1}} depicted in Fig. \ref{pickassemblemove}b, as shown in Fig. \ref{singvalues}. The transfer function between ${\mathbf{T}_{\mathrm{ext} / \mathcal{B}, G}}\{1\}$ and $\boldsymbol{\dot{{\omega}}}_{G}\{1\}$ is examined for the nominal system (${\mathbf{T}_{\mathrm{ext} / \mathcal{B}, G}}\{1\} \rightarrow \boldsymbol{\dot{{\omega}}}_{G}\{1\}$ channel). The plot is coherent with the properties of the spacecraft's flexible elements, in accordance with the definition of the modal participation factor matrices $\mathbf{L}_{P_{1}}^{\mathcal{A}}$ and $\mathbf{L}_{P_{2}}^{\mathcal{F}_3}$ \cite{Guy2014}. Indeed, the antiresonances occur at the frequencies of the cantilevered flexible modes corresponding to the structure being assembled and the solar array. The same figure also depicts the effect that the set of real parametric uncertainty $\boldsymbol{\Delta}_{\omega}$ has on the singular values of the same transfer function. Since the inverse linearized dynamic model of the whole system projected in $\mathcal{R}_{b}$ and the channel ${\mathbf{T}_{\mathrm{ext} / \mathcal{B}, G}}\{1\} \rightarrow \boldsymbol{\dot{{\omega}}}_{G}\{1\}$ are being analyzed, the static gain of the plot in Fig. \ref{singvalues} represents the inverse of the first moment of inertia $1/{J}^{\text{tot}}_{xx}$ measured at point $G$ and with respect to $\mathcal{R}_{b}$ for the moment {\Large \textcircled{\normalsize 1}} depicted in Fig. \ref{pickassemblemove}b.

\subsubsection{Baseline attitude controller}

Some of the challenges of an on-orbit assembly mission scenario include the control structure interactions between the flexible appendages and the AOCS, the time-varying inertial properties and flexible dynamics, and also the dynamic couplings. Prior to designing various cost functions to assign weights to the multiple edges in all potential node graphs, the open loop dynamics of the diverse arrays of models $\mathbf{H}^m(\mathrm{s})$, which represent these edges, are closed using a controller.

Initially, the AOCS design is divided into two different parts, namely the translational and the attitude. However, only attitude control design is addressed, since this paper's objective is to focus on the on-orbit assembly phase of the mission. A baseline attitude controller is tuned based on the inertial properties of the system, as follows: 

\begin{equation}
\mathbf{u}=\mathbf{K}_{\mathrm{att}}\left[\begin{array}{c}
\boldsymbol{\Theta}_G - \boldsymbol{\Theta}_G^{\mathrm{ref}}\\
 \boldsymbol{\omega}_G - \boldsymbol{\omega}_G^{\mathrm{ref}}
\end{array}\right] \quad \text{and} \quad
\mathbf{K}_{\mathrm{att}}=\left[\begin{array}{ll}\mathbf{k}_{\mathrm{att}} & \mathbf{c}_{\mathrm{att}}\end{array}\right]
\quad \text{with} \quad
\begin{cases}
  \mathbf{k}_{\mathrm{att}}=- \omega_{\mathrm{att}}^{2} \textbf{J}^{\text{tot}}_{G}\\
  \mathbf{c}_{\mathrm{att}}=- 2 \xi_{\mathrm{att}} \omega_{\mathrm{att}} \textbf{J}^{\text{tot}}_{G}
\end{cases}  
\label{control}
\end{equation}

\begin{figure}[!ht]
\centering
\includegraphics[width=1\textwidth]{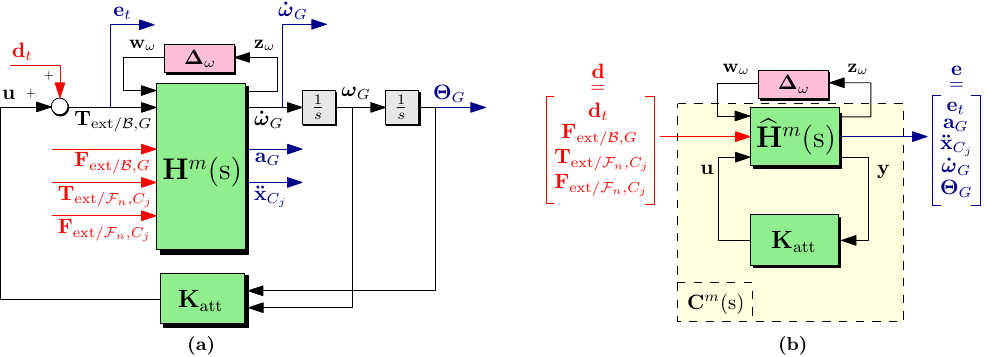}
\caption{(a) Closed-loop model. (b) Equivalent standard form of the interconnection.}
\label{controlMAR} 
\end{figure}

In Eq. \eqref{control}, $\mathbf{u}$ represents the torque control output and $\boldsymbol{\omega}_G$ is the inertial angular velocity of the spacecraft defined in $\mathcal{R}_{b}$ and obtained from integrating $\boldsymbol{\dot{\omega}}_G$. Similarly, the linearized \textsc{Euler} angles vector $\boldsymbol{{\Theta}}_G$, computed with respect to the inertial frame, is obtained from a double integration of $\boldsymbol{\dot{\omega}}_G$, assuming small angular rates (linear model). Moreover, $\textbf{J}^{\text{tot}}_{G}$ is the inertia tensor of the collection of all the body elements, measured with respect to $\mathcal{R}_{b}$. In this case, $\textbf{J}^{\text{tot}}_{G}$ is computed under the assumption that the flexible structure consists of $27$ tiles (as in illustration {\Large \textcircled{\normalsize 3}} in Fig. \ref{Diapositive1}) and that the robotic arm $\mathcal{R}_{1}$ is connected to the tile number $22$, a scenario which corresponds to the worst-case inertia in terms of rejection of disturbances that are caused by the multi-arm robot. Additionally, $\alpha^{\mathcal{R}_{\bullet}}_{k}=0$ \si{\radian} and the robot is holding the last tile to be assembled. The objective is to have a critically damped system that returns to rest slowly without oscillating when tracking the reference signals $\boldsymbol{\Theta}_G^{\mathrm{ref}}$ and $\boldsymbol{\omega}_G^{\mathrm{ref}}$ (both of which, in this case, are equal to $\mathbf{0}_{3\times 1}$). For this reason, $\xi_{\mathrm{att}}=1$ and $\omega_{\mathrm{att}}=0.01$ \si{\hertz} represent the desired controller's damping ratio and natural frequency, respectively. These values of $\xi_{\mathrm{att}}$ and $\omega_{\mathrm{att}}$ are merely a tuning guess which were chosen to avoid overshoot and to achieve a large settling time. The closed-loop model of the system is depicted in Fig. \ref{controlMAR}a.

Ultimately, for every edge $m$, the resulting closed-loop model is given by $\mathbf{C}^m(\mathrm{s})$, as can be observed in Fig. \ref{controlMAR}b, with $\mathbf{C}^m(\mathrm{s})=\mathcal{F}_l\left(\widehat{\mathbf{H}}^m(\mathrm{s}), \mathbf{K}_{\mathrm{att}}\right)$, where $\mathcal{F}_{l}(\cdot)$ represents the lower linear fractional transformation. Furthermore, the uncertain closed-loop model is given by $\widehat{\mathbf{C}}^m(\mathrm s,\boldsymbol{\Delta}_{\omega})=\mathcal{F}_u\left({\mathbf{C}^m}(\mathrm s), \boldsymbol{\Delta}_{\omega}\right)$, where $\mathcal{F}_{u}(\cdot)$ represents the upper linear fractional transformation. The goal is to utilize the knowledge of the singular values of these arrays of closed-loop dynamical systems, which characterize the system's behavior when the moving robot performs a trajectory, to compute the cost of executing these trajectories.

\subsubsection{Impact of a wrench disturbance generated by the multi-arm robot on the angular acceleration of the spacecraft}
\label{hinfhardconst}

In this case, the considered cost function computes the $\mathcal{H}_{\infty}$ norm (or peak gain) of the transfer matrix between a wrench that is applied by the robot to the FEM of $\mathcal{F}_n$ at the position where the multi-arm robot is connected to the structure, which is given by $\mathbf{W}_{\mathrm{ext}/\mathcal{F}_n,C_j}={\left[\begin{array}{c}\mathbf{F}_{\mathrm{ext}/\mathcal{F}_n,C_j}\\ \mathbf{T}_{\mathrm{ext}/\mathcal{F}_n,C_j}\end{array}\right]}$, and the angular acceleration $\boldsymbol{\dot{\omega}}_G$. It is important to note that this transfer function is chosen simply as an example. Many other transfer functions/matrices could be employed for the cost function calculation. In this case, the transfer $\mathbf{W}_{\mathrm{ext} / \mathcal{F}_n, C_j} \rightarrow \boldsymbol{\dot{\omega}}_G$ was selected because a wrench disturbance applied by the robot generates an angular acceleration on the spacecraft (see Fig. \ref{COMPLETEMODEL_mirror_fig}). The selection of the transfer function/matrix being analyzed naturally depends on the type of structure being assembled and also on the mission requirements. This cost function is computed for every edge $m$, as follows:

\begin{equation}
J^m_{{\mathcal{H}_{\infty}},\mathbf{W}\boldsymbol{\dot{\omega}}}=\sum_{l=1}^{2z} \left\|{\mathbf{C}_l^m(\mathrm{s})}_{\mathbf{W}_{\mathrm{ext} / \mathcal{F}_n, C_j} \rightarrow \boldsymbol{\dot{\omega}}_G}\right\|_{\infty} 
\label{costleverage}
\end{equation}

In Eq. \eqref{costleverage}, ${\mathbf{C}_l^m(\mathrm{s})}$ represents the dynamical system number $l$ of the array $\mathbf{C}^m(\mathrm{s})$, which is logically associated with the edge number $m$. Furthermore, it should be noted that the uncertainty block $\boldsymbol{\Delta}_{\omega}$ is considered nominal. After computing the cost function $J^m_{{\mathcal{H}_{\infty}},\mathbf{W}\boldsymbol{\dot{\omega}}}$ for all the edges $m$, the different node graphs are populated with the different costs for all the possible trajectories of the multi-arm robot during the assembly scenario. In this paper, $N=28$ is considered (see illustration {\Large \textcircled{\normalsize 3}} in Fig. \ref{Diapositive1}), indicating that there are $56$ possible node graphs for the assembly of the structure $\mathcal{F}_{28}$. Furthermore, the total number of edges in these $56$ node graphs is equal to $M=8672$. 

Let us now compare, for the structure $\mathcal{F}_{27}$, the difference in singular values between the weighted node graphs (using the cost function in Eq. \eqref{costleverage}) and unweighted node graphs when computing the full trajectory between the tiles number 2 and 27, considering the multi-arm robot is initially connected to the tile number 2 with the robotic arm $\mathcal{R}_1$. Additionally, another input required for the algorithm to function is specifying the robotic arm that should be connected to the final tile, which is in this case the tile number 27. For this particular trajectory computation, it is determined that the robotic arm $\mathcal{R}_2$ should be connected at the end of the trajectory. Furthermore, it is also assumed that $\delta=1$ (the multi-arm robot is carrying a tile).

For the unweighted node graphs, a Breadth-First computation \cite{cormen} that treats all edge weights as 1 is used to find the shortest path between the tiles number 2 and 27. For the weighted node graphs, the Dijkstra algorithm \cite{cormen} is used to compute the optimal path.

\begin{figure}[!ht]
\centering
\includegraphics[width=1\textwidth]{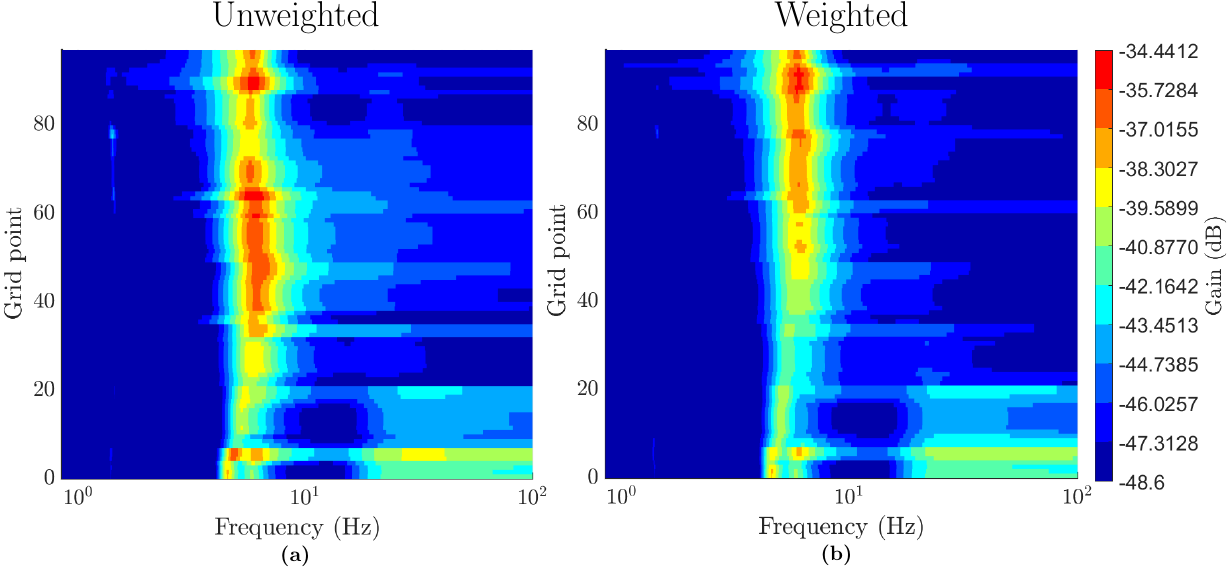}
\caption{Singular values of the channels $\mathbf{W}_{\mathrm{ext} / \mathcal{F}_{27}, C_j} \rightarrow \boldsymbol{\dot{\omega}}_G$ across a dense grid of frequencies and different grid points: (a) unweighted. (b) weighted trajectory using the cost function depicted in Eq. \eqref{costleverage}.}
\label{surf_pathopt} 
\end{figure}

\begin{figure}[!ht]
\centering
\includegraphics[width=1\textwidth]{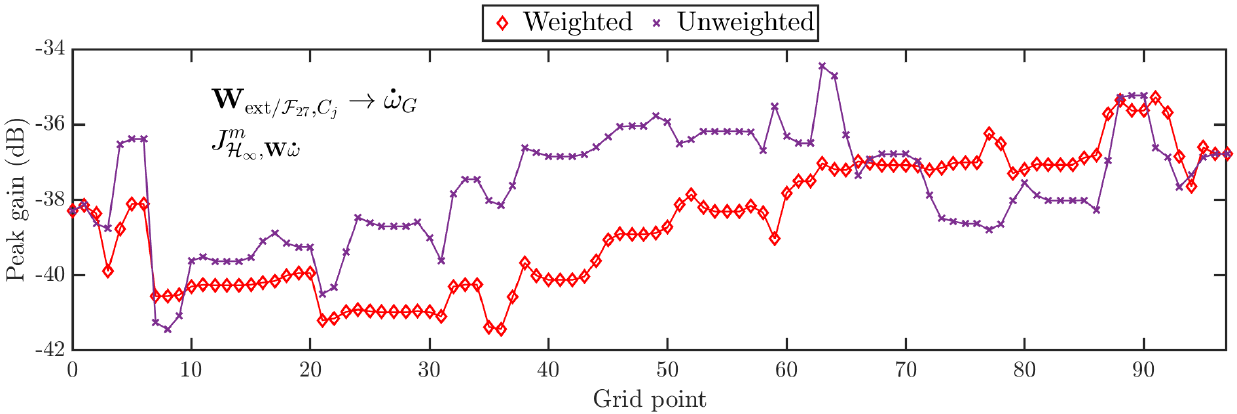}
\caption{Peak gains of the channels $\mathbf{W}_{\mathrm{ext} / \mathcal{F}_{27}, C_j} \rightarrow \boldsymbol{\dot{\omega}}_G$ for different grid points and trajectory computations.}
\label{hinf_grid} 
\end{figure}

\begin{figure}[!ht]
\centering
\includegraphics[width=1\textwidth]{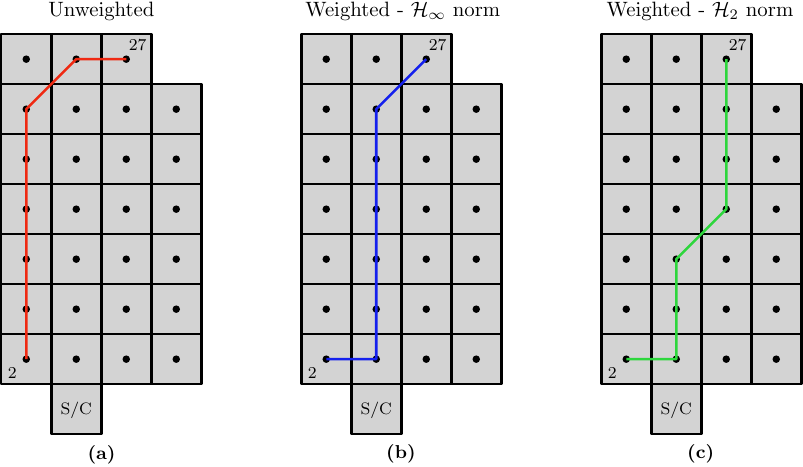}
\caption{Three different computed trajectories between the tiles number 2 and 27 for the multi-arm robot: (a) unweighted. (b) weighted considering the $\mathcal{H}_{\infty}$ cost function displayed in Eq. \eqref{costleverage}. (c)  weighted considering the $\mathcal{H}_{2}$ cost function displayed in Eq. \eqref{pointing}.}
\label{compare_trajectories} 
\end{figure}

Figure \ref{surf_pathopt} demonstrates that the peak gains of the singular values for the various grid points representing the entire trajectory are generally lower in Fig. \ref{surf_pathopt}b than those in Fig. \ref{surf_pathopt}a, which represents the singular values computation without an associated cost function. The peak gain values are more clearly depicted in Fig. \ref{hinf_grid}. This figure further illustrates that, for the majority of the grid points, the unweighted peak gain values are higher. To support this observation, the cumulative cost was calculated for both the unweighted and the weighted trajectories along all the grid points, revealing that the unweighted cumulative cost is approximately  $11.5 \%$ higher than the weighted cumulative cost. Furthermore, Figs. \ref{compare_trajectories}a and \ref{compare_trajectories}b illustrate the distinct trajectories chosen by the algorithm for the unweighted and weighted scenarios throughout the entire path between the tiles number 2 and 27. It is important to note that, since the objective is to minimize the impact of a wrench on the spacecraft's angular acceleration, the weighted algorithm tries to keep the multi-arm robot closer to the spacecraft's center of mass, effectively reducing the impact of this wrench on the behaviour of the spacecraft.

 In both the unweighted and weighted scenarios, the moving robot takes $7$ small trajectory sub-steps to reach its final destination (Figs. \ref{compare_trajectories}a and \ref{compare_trajectories}b). Furthermore, the dynamical system arrays were gridded over \(z=7\) grid points, meaning that for each edge \(m\), the array \(\mathbf{C}^m(\mathrm{s})\) contains \(2z=14\) dynamical systems. As a result, Fig. \ref{hinf_grid} displays 98 grid points for both the weighted and unweighted trajectories (grid points start at $0$).

 This algorithm can also be employed to optimize a trajectory using the same cost function while mitigating peak gains by applying a hard constraint. For instance, consider the following constraint: if the peak gain exceeds $x$ \si{\decibel}, the cost function is set to \(\infty\) for that particular grid point. This approach could help to avoid problematic peak gains, although it may require the multi-arm robot to take additional sub-steps to reach its final destination.

This type of algorithm can also be successively applied to all the different node graphs and linear models that describe the entire assembly scenario. The initial system configuration is assumed to be the one depicted in illustration {\Large \textcircled{\normalsize 1}} of Fig. \ref{Diapositive1}, but with $\alpha^{\mathcal{R}_{\bullet}}_{k}=0$ \si{\radian}. Additionally, the final computed trajectory corresponds to illustration {\Large \textcircled{\normalsize 3}} in Fig. \ref{Diapositive1}, where the robot is assembling the last tile in the structure.

\begin{figure}[!ht]
\centering
\includegraphics[width=1\textwidth]{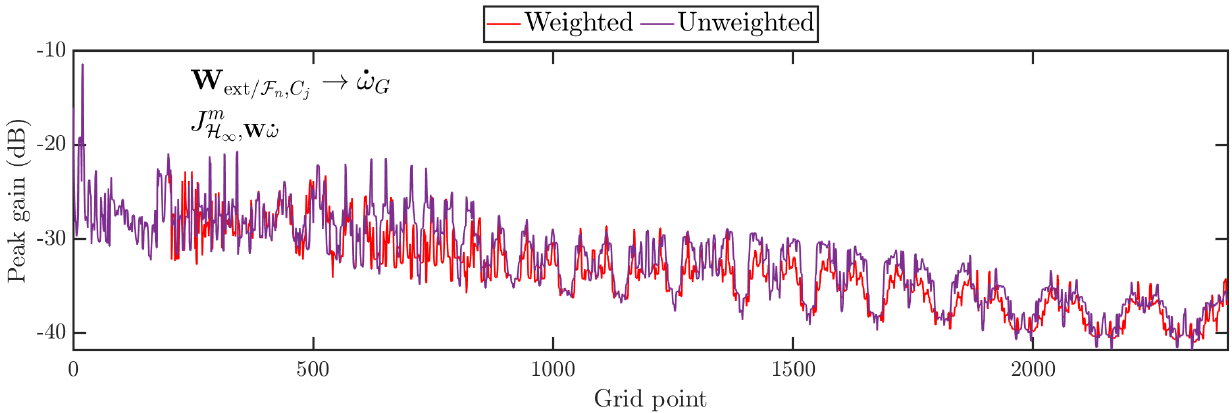}
\caption{Peak gains of the channels $\mathbf{W}_{\mathrm{ext} / \mathcal{F}_{n}, C_j} \rightarrow \boldsymbol{\dot{\omega}}_G$ for different grid points and trajectory computations.}
\label{hinf_grid_full} 
\end{figure}

The results, shown in Fig. \ref{hinf_grid_full}, are very similar to those obtained when the algorithm is applied to the trajectory computation of the multi-arm robot when considering solely the FEM of $\mathcal{F}_{27}$. In fact, calculating the cumulative cost of performing the entire assembly indicates that the cumulative unweighted cost over all the grid points is approximately $12\%$ higher than the cumulative weighted cost. Minimizing these peak gains is crucial for an on-orbit assembly scenario, as it reduces the impact on the spacecraft's angular accelerations when the flexible structure is hit with disturbances at problematic frequencies. This reduction in effect also alleviates the strain on the controller, thereby decreasing the control effort that is required to control the system. However, it should be noted that the focus here is on the peak gain, although it is worth mentioning that this peak might not be excited. Nevertheless, this approach ensures coverage for the worst-case scenario. A different characterization of this cost function could be achieved using the $\mathcal{H}_2$ norm, as proposed in section \ref{sech2}, especially if the goal is to minimize the energy that is transferred to the central hub.

It is again important to note that any transfer function can be employed to compute cost functions within this algorithm. While the results appear intuitive (a wrench generates an angular acceleration) for the current structure and cost function, analyzing different performance and stability metrics or even other more complex structures might be less intuitive. Nevertheless, this algorithm is fully generalized for any structure composed of modular tiles, where the only necessary changes involve the modeling of the system and the computation of the node graphs.

Ultimately, a summary of all the steps that are required to apply the path optimization algorithm is also depicted in Algorithm \ref{alg:cap}.

\begin{algorithm}
\caption{Main steps of the path optimization algorithm}\label{alg:cap}
{Compute the family of $2N(N+1)$ models that describe the assembly scenario (assembly order of the structure must be predefined)$;$} \Comment{see Fig. \ref{COMPLETEMODEL_mirror_block}} 

{Define the adjacency matrices of the node graphs$;$}  

{Build the node graphs for picking up and assembling a tile$;$} \Comment{see Fig. \ref{node_graph}}

{Grid the different dynamic models by replacing $\boldsymbol{\Delta}_{\mathcal{R}_1}$, $\boldsymbol{\Delta}_{\mathcal{R}_2}$, $\boldsymbol{\Delta}_{\mathcal{R}_3}$ and $\delta$ according to the different actions the multi-arm robot can perform: picking up a tile, walking and assembling a tile$;$} \Comment{see Fig. \ref{pickassemblemove}}

{Compute the closed-loop model for every edge $m$$;$} \Comment{see Fig. \ref{controlMAR}}

{Compute a cost function based on certain stability or performance requirements$;$} \Comment{see Eq. \eqref{costleverage} as example}

{Use this cost function to weight every edge $m$ and generate the optimal trajectory for the multi-arm robot$;$} 

\end{algorithm}

\subsubsection{Impact of a wrench disturbance generated by the multi-arm robot on the pointing performance of the spacecraft}

\label{sech2}

A different performance cost function is considered, which computes the $\mathcal{H}_{2}$ norm of the transfer function between the closed-loop external wrench disturbances $\mathbf{W}_{\mathrm{ext}/\mathcal{F}_n,C_j}$ and the \textsc{Euler} angles $\boldsymbol{{\Theta}}_G$. In this case, the $\mathcal{H}_{2}$ norm was selected because it measures the steady-state covariance (or power) of the output response $\boldsymbol{{\Theta}}_G=\mathbf{C}_l^m(\mathrm{s})\mathbf{W}_{\mathrm{ext}/\mathcal{F}_n,C_j}$ to unit white noise inputs (the uncertainty block $\boldsymbol{\Delta}_{\omega}$ is considered nominal), therefore assessing the pointing performance of the spacecraft when hit by disturbances. This cost function is computed for every edge $m$, as follows:

\begin{equation}
J^m_{{\mathcal{H}_{2}},\mathbf{W}\boldsymbol{\Theta}}=\sum_{l=1}^{2z} \left\|{\mathbf{C}_l^m(\mathrm{s})}_{\mathbf{W}_{\mathrm{ext}/\mathcal{F}_n,C_j} \rightarrow \boldsymbol{{\Theta}}_G}\right\|_{2} 
\label{pointing}
\end{equation}

\begin{figure}[!ht]
\centering
\includegraphics[width=1\textwidth]{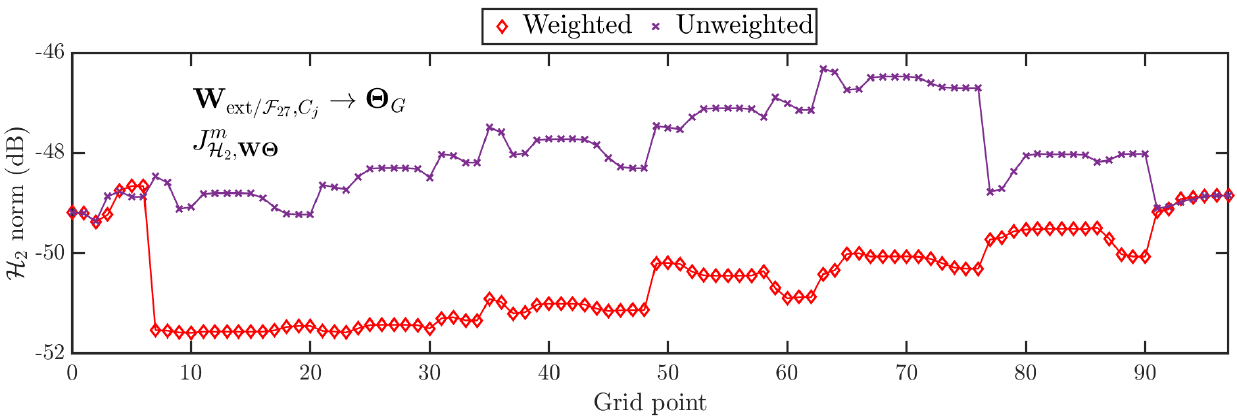}
\caption{$\mathcal{H}_{2}$ norm of the channels $\mathbf{W}_{\mathrm{ext}/\mathcal{F}_{27},C_j} \rightarrow \boldsymbol{{\Theta}}_G$ for different grid points and trajectory computations.}
\label{hinf_grid_h2} 
\end{figure}

Here, the comparison of singular values is also between the weighted and unweighted node graphs for the structure $\mathcal{F}_{27}$ when computing the robot's full trajectory between tiles number 2 and 27, assuming the robotic arm $\mathcal{R}_1$ is connected at the beginning and the robotic arm $\mathcal{R}_2$ at the end of the trajectory, with $\delta=1$. The objective is to limit the closed-loop system's energy to reduce the attitude impact of an external wrench disturbance applied on the FEM of $\mathcal{F}_{27}$. As shown in Fig. \ref{hinf_grid_h2}, the weighted energy computation remains below the unweighted one for most grid points along the trajectory. Additionally, the cumulative cost calculation for the entire trajectories shows that the unweighted cost is approximately $32.7\%$ higher than the weighted cost, showing once again the effectiveness of the proposed optimization algorithm. Additionally, Fig. \ref{compare_trajectories}c illustrates the trajectory selected by the algorithm for the weighted case using the cost function provided in Eq. \eqref{pointing}. A demonstrative video of this path optimization algorithm is available at \url{https://www.youtube.com/watch?v=VkGZAHeFZ4o}. This video shows the trajectories depicted in Fig. \ref{compare_trajectories}, along with the associated node graphs and singular values of the respective transfer functions.

\subsubsection{Stability margins}

A different cost function is now analyzed, which computes the $\mathcal{H}_{\infty}$ norm (or peak gain) of the input sensitivity transfer function between the closed-loop external torque disturbances $\mathbf{d}_t$ which are applied to the spacecraft and the torques $\mathbf{e}_t$ that are applied to the spacecraft by the actuators (the uncertainty block $\boldsymbol{\Delta}_{\omega}$ is considered nominal), as depicted in Fig. \ref{controlMAR}. This cost function is computed for every edge $m$, as follows:

\begin{equation}
J^m_{{\mathcal{H}_{\infty}},\mathbf{d}\mathbf{e}}=\sum_{l=1}^{2z} \left\|{\mathbf{C}_l^m(\mathrm{s})}_{\mathbf{d}_{{t} } \rightarrow \mathbf{{e}}_t}\right\|_{\infty} 
\label{costinputsensitivity}
\end{equation}

\begin{figure}[!ht]
\centering
\includegraphics[width=1\textwidth]{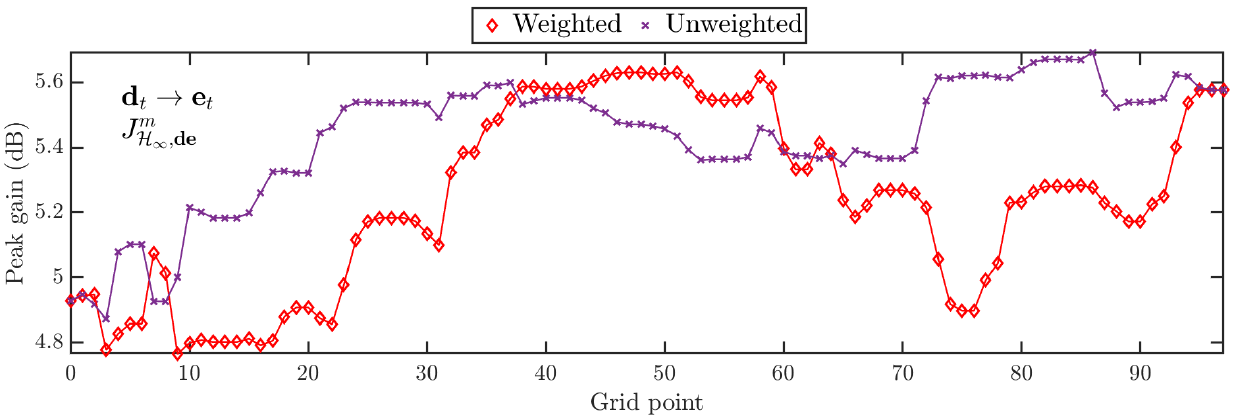}
\caption{Peak gains of the channels $\mathbf{d}_{{t} } \rightarrow \mathbf{{e}}_t$ for different grid points and trajectory computations.}
\label{hinf_grid_inputsensit} 
\end{figure}

Let us now compare the difference in singular values between the weighted and unweighted node graphs for the structure $\mathcal{F}_{27}$, when computing the full trajectory of the robot between the tiles number 2 and 27. This comparison assumes the multi-arm robot is initially connected to the tile number 2 with the robotic arm $\mathcal{R}_1$. Furthermore, it is once again determined that the robotic arm $\mathcal{R}_2$ should be connected at the end of the trajectory, with $\delta=1$. The peak gains obtained across all the different grid points for the entire trajectory are shown in Fig. \ref{hinf_grid_inputsensit}.

In this instance, the algorithm achieves a trajectory that meets the cost function requirements. The cumulative cost calculation for the entire trajectories shows that the unweighted cost is approximately $2.2\%$ higher than the weighted cost. Although the improvements are small compared to the unweighted shortest path computation, this is directly related to the system's physics and dynamics. If the improvements are not larger, it indicates a true physical constraint, meaning there were no other paths for the robot that could enhance the analyzed metrics. However, by reducing the peak gain of the input sensitivity function even by a small fraction, the system's modulus, gain and phase margins along the three axes are automatically increased.

\subsubsection{Worst-case robustness analysis}

Robustness analysis involves determining whether the designed closed-loop system maintains specified stability margins and performance levels when subjected to modeling uncertainties. Let us now introduce the structured singular value function $\mu_{\delta}(\cdot)$ \cite{Doyle1993}, which provides very precise information about the magnitude of uncertainty which is needed to destabilize the loop at any frequency \cite{Preda2020}. If the nominal system (i.e., the block $\mathbf{C}^m_{{\mathbf{d}} \rightarrow {\mathbf{e}}}$) shown in Figure \ref{controlMAR} is stable, then the stability of this loop is conditioned by the existence of $\left(\textbf{I}-\mathbf{C}^m_{{\mathbf{w}_{\omega}} \rightarrow {\mathbf{z}_{\omega}}} \boldsymbol{\Delta}\right)^{-1}$. This is assessed by evaluating for different frequencies $\omega_{\mu} \in \mathbb{R}$ the structured singular value $\mu_{\delta}\left(\mathbf{C}^m_{{\mathbf{w}_{\omega}} \rightarrow {\mathbf{z}_{\omega}}}(j \omega_{\mu})\right)$, with $\mu_{\delta}(\cdot)$ being defined for a complex matrix $\mathbf{C}^m \in \mathbb{C}^{a \times b}$ and a set of uncertainties $\boldsymbol{\Delta} \in \Delta \subset \mathbb{R} \mathbb{H}_{\infty}^{b \times a}$, as follows:

\begin{equation}
\mu_{\delta}(\mathbf{C}^m)=\frac{1}{\min \{\bar{\sigma}(\boldsymbol{\Delta}): \boldsymbol{\Delta} \in {\Delta}, \operatorname{det}(\textbf{I}-\mathbf{C}^m \boldsymbol{\Delta})=0\}}
\label{mu}
\end{equation}

where $\mathbb{C}^{a \times b}$ is the set of $a$-by-$b$ complex matrices and the set $\mathbb{R} \mathbb{H}_{\infty}^{b \times a}$ describes the set of finite gain transfer matrices with $b$ outputs and $a$ inputs. For $\mathcal{G} \in \mathbb{R} \mathbb{H}_{\infty}^{b \times a}$, the value $\|\mathcal{G}\|_{\infty}$ represents the $\mathcal{L}_{2}$ system gain. If no $\boldsymbol{\Delta} \in {\Delta}$ makes $\textbf{I}-\mathbf{C}^m \boldsymbol{\Delta}$ singular, then $\mu_{\delta}(\mathbf{C}^m):=0$. Following this definition and under the assumption that the nominal system $\mathbf{C}^m_{{\mathbf{d}} \rightarrow {\mathbf{e}}}$ is stable, then $\mathcal{F}_{u}(\mathbf{C}^m, \boldsymbol{\Delta})$ is stable $\forall \boldsymbol{\Delta} \in {\Delta}$, $\bar{\sigma}(\boldsymbol{\Delta})<\nu$ if and only if: $\mu_{\delta}\left(\mathbf{C}^m_{\mathbf{w}_{\omega} \rightarrow \mathbf{z}_{\omega}}(j \omega_{\mu})\right)<1 / \nu , \forall \omega_{\mu} \in \mathbb{R}$. In this context, $\mu_{\delta}$ gives a measure of the smallest structured uncertainty $\boldsymbol{\Delta}$ that causes closed-loop instability for any frequency $\omega_{\mu} \in$ $\mathbb{R}$. Moreover, the $\mathcal{L}_{2}$ gain of this destabilizing perturbation is exactly $1 / \mu_{\delta}$. This fact will be used to evaluate the stability margins of the loop with respect to $\boldsymbol{\Delta}_{\omega}$. However, due to its non-convex character, $\mu_{\delta}$ can be difficult to compute exactly. For this reason, some very efficient algorithms \cite{balas2007robust} have been developed in order to estimate the bounds of $\mu_{\delta}$. 

In this case, the cost function under consideration calculates the upper bounds of the structured singular value function $\mu_{\delta}$ for the uncertainty set $\boldsymbol{\Delta}_{\omega}$, thereby evaluating the robust stability of ${\mathbf{C}_l^m(\mathrm{s})}$ for each edge $m$, as follows:

\begin{equation}
J^m_{\mu,\mathbf{w\mathbf{z}}}=\sum_{l=1}^{2z} \mu_{\delta}({\mathbf{C}_l^m(\mathrm{s})_{{\mathbf{w}_{\omega}} \rightarrow {\mathbf{z}_{\omega}}}}) 
\label{mucost}
\end{equation}

\begin{figure}[!ht]
\centering
\includegraphics[width=1\textwidth]{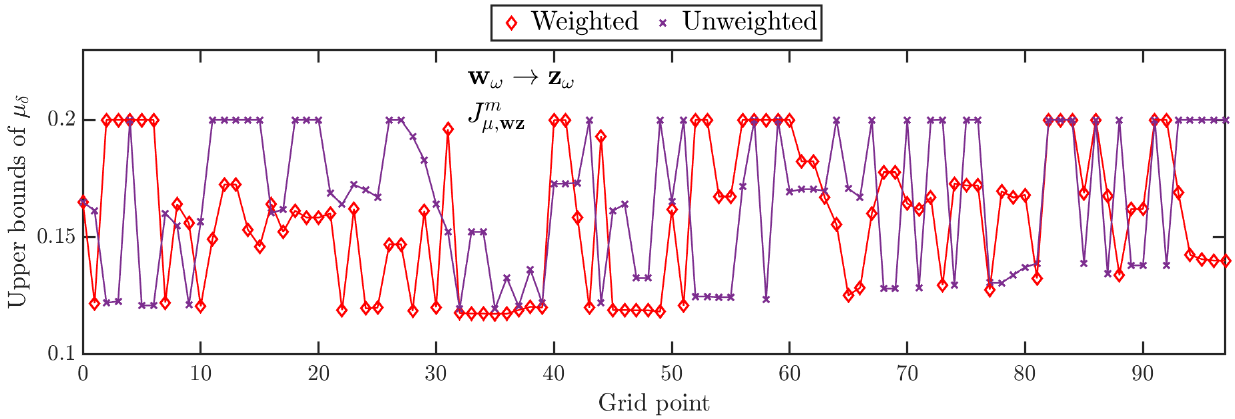}
\caption{Upper bounds of $\mu_\delta$ for different grid points and trajectory computations.}
\label{munormalpartial_stability} 
\end{figure}

In this case, the comparison of upper bounds of $\mu_{\delta}$ is also between the weighted and unweighted node graphs for the structure $\mathcal{F}_{27}$ when computing the robot's full trajectory between tiles number 2 and 27, following the same assumptions as before. Fig. \ref{munormalpartial_stability} demonstrates that the upper bounds of $\mu_{\delta}$ for the various grid points representing the entire trajectory are generally lower for the weighted computation. Additionally, the cumulative cost calculation for the entire trajectories shows that the unweighted cost is approximately $4.3\%$ higher than the weighted cost. Moreover, the maximum upper bound of $\mu_{\delta}$ for the weighted case across all grid points of the computed trajectory is equal to 0.2. This implies that the loop can tolerate a 400\% increase (for the particular grid point corresponding to the maximum upper bound) in the uncertainty $\boldsymbol{\Delta}_{\omega}$ while still maintaining stability, demonstrating significant robust stability with respect to the uncertainty block $\boldsymbol{\Delta}_{\omega}$. Logically, uncertainty could also be considered for other parameters, such as the mass and inertia of the rigid hub $\mathcal{B}$. In this case, the structured singular value analysis could be extended to an uncertainty block consisting of multiple subsets of uncertainties. Ultimately, the impact of the uncertainty set $\boldsymbol{\Delta}_{\omega}$ could also be assessed on different performance indicators using structured singular value computations.

For this specific cost function, due to the physical interpretation of the structured singular value function $\mu_{\delta}$, it would be particularly interesting to consider an uncertainty block that drives the upper bounds of $\mu_{\delta}$ close to 1. In this scenario, the algorithm could be highly valuable, especially when combined with a hard constraint (as explained in section \ref{hinfhardconst}), as it could prevent situations where $\mu_{\delta} > 1$, which would indicate that the system cannot account for the entire uncertainty block being considered.

\section{Conclusion}

In conclusion, this paper presents a comprehensive framework for modeling and optimizing the linear dynamic behavior of on-orbit assembly scenarios involving large flexible structures using a three-legged robot. The developed models effectively capture the dynamic interactions and parameterization effects of the assembly process, which is marked by time-varying flexible dynamics and inertial properties. Additionally, the paper introduces an innovative method for optimizing the path of multi-arm robots during the assembly process. This method incorporates various cost functions defined by performance and stability metrics, ensuring the efficient and precise movement of robots across the flexible structure. The combination of these modeling and optimization techniques provides a robust framework for understanding and improving the dynamics of on-orbit assembly. The results presented validate the effectiveness of the proposed methodology, showcasing its potential to significantly enhance the predictability, control and efficiency of on-orbit assembly operations. This work paves the way for more advanced robotic assembly techniques in space, ultimately contributing to the advancement of space engineering and the realization of complex space structures. Ultimately, it should be noted that the methodology described in this paper is based on the assumption that the system experiences minor deflections and maintains mostly linear dynamics. This assumption is applicable to a broad range of space-related applications, as spacecraft typically avoid large deflections and nonlinearities in structural dynamics by design.

\begin{table} [!ht]

	\caption{Spacecraft mechanical data. \textbf{Nomenclature}: MoI (Moment of Inertia); PoI (Product of Inertia); CoM (Center of Mass).}
	\label{tab:Sat_prop}	
	\centering
	\resizebox{\textwidth}{!}{
	\begin{tabular}{p{1.5cm} l l r }
		\toprule
		& \textbf{Parameter} & \textbf{Description}  & \textbf{Value and Uncertainty} \\
		\midrule
		\multirow{12}{1.5cm}{\centering Rigid hub $\mathcal{B}$}
		& ${\mathbf{GP_1}}$ & distance vector between $G$ and $P_{1}$ written in $\mathcal{R}_{b}=\left(G; \mathbf{x}_{b}, \mathbf{y}_{b}, \mathbf{z}_{b}\right)$ & $[0,\, - 0.5,\, 0]$ \si{\meter} \\
  & ${\mathbf{GP_2}}$ & distance vector between $G$ and $P_{2}$ written in $\mathcal{R}_{b}$ & $[-0.5,\, 0.5,\,  0.7125]$ \si{\meter} \\
  & ${\mathbf{GP_3}}$ & distance vector between $G$ and $P_{3}$ written in $\mathcal{R}_{b}$ & $[0.5,\, 0,\, 0.7125]$ \si{\meter} \\
	
		&  $m^{{\mathcal{B}}}$ & mass of $\mathcal{B}$ & $166\ \si{\kilogram}$ \\
		& ${\left[\begin{array}{ccc}
{J}^{{\mathcal{B}}}_{xx} & {J}^{{\mathcal{B}}}_{xy} & {J}^{{\mathcal{B}}}_{xz} \\
 & {J}^{{\mathcal{B}}}_{yy} & {J}^{{\mathcal{B}}}_{yz} \\
 & & {J}^{{\mathcal{B}}}_{zz} 
\end{array}\right]}$ & inertia of $\mathcal{\mathcal{B}}$ at $G$ written in $\mathcal{R}_{b}$& $\begin{bmatrix}
		21.6256 & 3.84 & 0 \\
		 &  15.6256 & 0 \\
		 &  &  30.6738 \\
		\end{bmatrix}\, \si{\kilogram\square\meter} $ \\ 

  &  $\mathbf{P}_{\mathcal{R}_{a}/\mathcal{R}_{b} }$ & change of frame DCM between $\mathcal{R}_{a}=\left(P_1; \mathbf{x}_{a}, \mathbf{y}_{a}, \mathbf{z}_{a}\right)$ and $\mathcal{R}_{b}$ & $\begin{bmatrix}
		-1 & 0 & 0 \\
		 0 &  -1 & 0 \\
		0 & 0 &  1 \\
		\end{bmatrix} $ \\

  &  $\mathbf{P}_{\mathcal{R}_{s}/\mathcal{R}_{b} }$ & change of frame DCM between $\mathcal{R}_{s}=\left(P_3; \mathbf{x}_{s}, \mathbf{y}_{s}, \mathbf{z}_{s}\right)$ and $\mathcal{R}_{b}$ & $\mathbf{I}_3$ \\
  
  &  $\mathbf{P}_{\mathcal{R}_{f_n}/\mathcal{R}_{b} }$ & change of frame DCM between $\mathcal{R}_{f_n}=\left(P_2; \mathbf{x}_{f_n}, \mathbf{y}_{f_n}, \mathbf{z}_{f_n}\right)$ and $\mathcal{R}_{b}$ & $\mathbf{I}_3$ \\
  
  \midrule
		
		\multirow{13}{1.5cm}{\centering Solar array ${\mathcal{A}}$}
		& $\mathbf{P_{1}  S_{1}}$ & distance vector between $P_{1}$ and $S_{1}$ written in $\mathcal{R}_{a}$ & $[0,\, 1.0934,\, 0.0014] \ \si{\meter}$ \\
		& $m^{\mathcal{A}}$ & mass of $\mathcal{A}$ & $88.93\ \si{\kilogram}$ \\
		& ${\left[\begin{array}{ccc}
{J}^{{\mathcal{A}}}_{xx} & {J}^{{\mathcal{A}}}_{xy} & {J}^{{\mathcal{A}}}_{xz} \\
 & {J}^{{\mathcal{A}}}_{yy} & {J}^{{\mathcal{A}}}_{yz} \\
 & & {J}^{{\mathcal{A}}}_{zz} 
\end{array}\right]}$ & inertia of $\mathcal{A}$ at $S_{1}$ written in $\mathcal{R}_{a}$ 
		& $\begin{bmatrix}
		33.0918 & 0 & 0 \\
		&  7.3819   & -0.0002 \\
		&  &  40.4578 \\
		\end{bmatrix}\, \si{\kilogram\square\meter} $ \\
		& $[\omega_{1}^{\mathcal{A}},\omega_{2}^{\mathcal{A}} ]$ & flexible modes' frequencies & $[1.2850 \pm 20\%,\ 6.5896]\ \si{\hertz}$ \\ 
		& $[\xi_{1}^{\mathcal{A}},\xi_{2}^{\mathcal{A}}]$ & flexible modes' damping & $0.01$ \\ 
		& $\textbf{L}_{P_{1}}^{\mathcal{A}}$ & modal participation factors & 
		$\begin{bmatrix}
	    -0.0007  & -0.0078  &  7.8872 &  11.7690  &  0.0005  &  0.0010 \\
        -7.9401  &  0   & 0.0007  & -0.0008  &  0.1089 &  12.1014 \\
        -0.3604 &   0  &  0.0006  &  0.0017 &  -2.6631 &   0.5399 \\
        0.0019 &  -0.0066   & 3.9818 &    0.9098 &  -0.0007 &  -0.0033 \\
        0.0272  &  0.0003  & -0.0145 &  -0.0019  &  0.4907  & -0.0221 \\
        -0.0010  &  0.0357  & -2.2185  & -0.2320  & -0.0029 &   0.0012 \\
		\end{bmatrix}$
		\\ \midrule

  		\multirow{1}{1.5cm}{\centering Stack of tiles ${\mathcal{S}}$}
		& $\mathbf{P_{3}  C_{0}}$ & distance vector between $P_{3}$ and $C_{0}$ written in $\mathcal{R}_{s}$ & $[0.5,\, 0,\, 0] \ \si{\meter}$ \\
		
		
		\\ \midrule

\multirow{18}{1.5cm}{\centering Flexible structure ${\mathcal{F}}_{1}$}

& $\mathbf{P_{2}  C_{1}}$ & distance vector between $P_{2}$ and $C_{1}$ written in $\mathcal{R}_{f_{1}}$ & $[0.5,\, 0.5,\, 0] \ \si{\meter}$ \\
		
		& $m^{\mathcal{F}_{1}}$ & mass of $\mathcal{F}_{1}$ & $6.0423\ \si{\kilogram}$ \\
		& ${\left[\begin{array}{ccc}
{J}^{{\mathcal{F}_{1}}}_{xx} & {J}^{{\mathcal{F}_{1}}}_{xy} & {J}^{{\mathcal{F}_{1}}}_{xz} \\
 & {J}^{{\mathcal{F}_{1}}}_{yy} & {J}^{{\mathcal{F}_{1}}}_{yz} \\
 & & {J}^{{\mathcal{F}_{1}}}_{zz} 
\end{array}\right]}$ & inertia of $\mathcal{F}_{1}$ at $P_{2}$ written in $\mathcal{R}_{f_{1}}$ 
		& $\begin{bmatrix}
		2.0147 & 1.5106 & 0 \\
		&  2.0147   & 0 \\
		&  &  4.0282 \\
		\end{bmatrix}\, \si{\kilogram\square\meter} $ \\
		& $[\omega_{1}^{\mathcal{F}_{1}},\omega_{2}^{\mathcal{F}_{1}},\omega_{3}^{\mathcal{F}_{1}} ]$ & flexible modes' frequencies & $[30.7169,\  35.1732,\ 50.2930]\ \si{\hertz}$ \\ 
		& $[\xi_{1}^{\mathcal{F}_{1}},\xi_{2}^{\mathcal{F}_{1}},\xi_{3}^{\mathcal{F}_{1}}]$ & flexible modes' damping & $0.005$ \\ 
		& $\textbf{L}_{P_{2}}^{\mathcal{F}_{1}}$ & modal participation factors & 
		$\begin{bmatrix}
	    -0.2004  & 0.2004  & 0 &   0  &  0  &  0.6902 \\
       0  &  0   & -0.3290  & -0.5375  &  0.5375 &  0 \\
        0 &   0  &  0  &  0.9139 &  0.9139 &   0 \\
        
		\end{bmatrix}$ \\

  & $\mathbf{\Phi}_{C_{1}}^{\mathcal{F}_{1}}$ & projection matrix of the modal shapes & 
		$\begin{bmatrix}
	    -1.2677  & 0  &  0 \\
        1.2677 &  0   & 0  \\
        0 &   -2.0936  &  0   \\
        0 &  -1.3080   & 0.5019 \\
       0  &  -0.7856  & 0.5019\\
        0.6827  &  0  & 0  \\
		\end{bmatrix}$
		\\

  \midrule
  
\multirow{18}{1.5cm}{\centering Flexible structure ${\mathcal{F}}_{26}$}

& $\mathbf{P_{2}  C_{25}}$ & distance vector between $P_{2}$ and $C_{25}$ written in $\mathcal{R}_{f_{26}}$ & $[-0.5,\, 6.5,\, 0] \ \si{\meter}$ \\
		
		& $m^{\mathcal{F}_{26}}$ & mass of $\mathcal{F}_{26}$ & $157.1\ \si{\kilogram}$ \\
		& ${\left[\begin{array}{ccc}
{J}^{{\mathcal{F}_{26}}}_{xx} & {J}^{{\mathcal{F}_{26}}}_{xy} & {J}^{{\mathcal{F}_{26}}}_{xz} \\
 & {J}^{{\mathcal{F}_{26}}}_{yy} & {J}^{{\mathcal{F}_{26}}}_{yz} \\
 & & {J}^{{\mathcal{F}_{26}}}_{zz} 
\end{array}\right]}$ & inertia of $\mathcal{F}_{26}$ at $P_{2}$ written in $\mathcal{R}_{f_{26}}$ 
		& $\begin{bmatrix}
		2251.79 & 78.55 & 0 \\
		&  209.48   & 0 \\
		&  &  2461.24 \\
		\end{bmatrix}\, \si{\kilogram\square\meter} $ \\
		& $[\omega_{1}^{\mathcal{F}_{26}},\omega_{2}^{\mathcal{F}_{26}},\omega_{3}^{\mathcal{F}_{26}} ]$ & flexible modes' frequencies & $[0.9120,\ 2.1,\ 2.99]\ \si{\hertz}$ \\ 
		& $[\xi_{1}^{\mathcal{F}_{26}},\xi_{2}^{\mathcal{F}_{26}},\xi_{3}^{\mathcal{F}_{26}}]$ & flexible modes' damping & $0.005$ \\ 
		& $\textbf{L}_{P_{2}}^{\mathcal{F}_{26}}$ & modal participation factors & 
		$\begin{bmatrix}
	    0  & 0  & -0.1427 &   -0.0276  &  0.0039  &  0 \\
       0.1289  &  0.0112   & 0  & 0  &  0 &  -0.0203 \\
        0 &   0  &  0.1352  &  0.0262 &  0.0680 &   0 \\
        
		\end{bmatrix}$ \\

  & $\mathbf{\Phi}_{C_{25}}^{\mathcal{F}_{26}}$ & projection matrix of the modal shapes & 
		$\begin{bmatrix}
	    0  & 9.9969  &  0 \\
        0  &  -2.7773   & 0  \\
        -10.3108 &   0  &  -2.9745   \\
        -47.2211 &  0   & 0.0796 \\
       -1.0374  &  0  & 12.7970\\
        0  &  -48.1478  & 0  \\
		\end{bmatrix}$
		\\

  &  $\mathbf{P}_{{\mathcal{R}_{{l^{\mathcal{R}_{1/2}}_0}}}/{\mathcal{R}_{{f_n}}}}$ & change of frame DCM between ${\mathcal{R}_{{l^{\mathcal{R}_{1/2}}_0}}}$ and ${\mathcal{R}_{{f_n}}}$ & $\mathbf{I}_3$ \\
  \midrule

	\end{tabular}
	
}
\end{table}

\begin{table} [!ht]

	\caption{Spacecraft mechanical data. \textbf{Nomenclature}: MoI (Moment of Inertia); PoI (Product of Inertia); CoM (Center of Mass).}
	\label{tab:Sat_prop2}	
	\centering
	\resizebox{\textwidth}{!}{
	\begin{tabular}{p{1.5cm} l l r }
		\toprule
		& \textbf{Parameter} & \textbf{Description}  & \textbf{Value and Uncertainty} \\
		\midrule

		\multirow{17}{1.5cm}{\centering Robotic arms $\mathcal{R}_{\bullet}$}
		& $m^{\mathcal{L}_{\bullet}^{\mathcal{R}_{\bullet}}}$ & mass of $\mathcal{L}_{\bullet}^{\mathcal{R}_{\bullet}}$ & $[5,\, 5,\, 10,\, 5,\, 10,\, 5]\ \si{\kilogram}$ \\
		& $x_{\mathcal{L}_{\bullet}^{\mathcal{R}_{\bullet}}}$ & x-coordinate of the CoM of $\mathcal{L}_{\bullet}^{\mathcal{R}_{\bullet}}$ written in $\mathcal{R}_{l^{\mathcal{R}_{\bullet}}_{\bullet}}$ & $[0,\, 0,\, -0.1062,\, 0,\, -0.1031,\, 0]\ \si{\meter}$ \\
		& $y_{\mathcal{L}_{\bullet}^{\mathcal{R}_{\bullet}}}$ & y-coordinate of the CoM of $\mathcal{L}_{\bullet}^{\mathcal{R}_{\bullet}}$ written in $\mathcal{R}_{l^{\mathcal{R}_{\bullet}}_{\bullet}}$ & $[0,\, 0,\, 0,\, 0,\, 0,\, 0]\ \si{\meter}$ \\
		& $z_{\mathcal{L}_{\bullet}^{\mathcal{R}_{\bullet}}}$ & z-coordinate of the CoM of $\mathcal{L}_{\bullet}^{\mathcal{R}_{\bullet}}$ written in $\mathcal{R}_{l^{\mathcal{R}_{\bullet}}_{\bullet}}$ & $[0.0625,\, 0.05,\, 0,\, 0.0810,\, 0,\, 0.0810]\ \si{\meter}$ \\ 
		& $J_{xx}^{{\mathcal{L}}_{\bullet}^{\mathcal{R}_{\bullet}}}$ & first MoI of $\mathcal{L}_{\bullet}^{\mathcal{R}_{\bullet}}$ at the CoM of $\mathcal{L}_{\bullet}^{\mathcal{R}_{\bullet}}$ written in $\mathcal{R}_{l^{\mathcal{R}_{\bullet}}_{\bullet}}$ & $[0.2,\, 0.2,\, 0.4,\, 0.2,\, 0.4,\, 0.2]\ \si{\kilogram\square\meter}$ \\
		& $J_{yy}^{{\mathcal{L}}_{\bullet}^{\mathcal{R}_{\bullet}}}$ & second MoI of $\mathcal{L}_{\bullet}^{\mathcal{R}_{\bullet}}$ at the CoM of $\mathcal{L}_{\bullet}^{\mathcal{R}_{\bullet}}$ written in $\mathcal{R}_{l^{\mathcal{R}_{\bullet}}_{\bullet}}$ & $[0.2,\, 0.2,\, 0.4,\, 0.2,\, 0.4,\, 0.2]\ \si{\kilogram\square\meter}$ \\		
		& $J_{zz}^{{\mathcal{L}}_{\bullet}^{\mathcal{R}_{\bullet}}}$ & third MoI of $\mathcal{L}_{\bullet}^{\mathcal{R}_{\bullet}}$ at the CoM of $\mathcal{L}_{\bullet}^{\mathcal{R}_{\bullet}}$ written in $\mathcal{R}_{l^{\mathcal{R}_{\bullet}}_{\bullet}}$ & $[0.2,\, 0.2,\, 0.4,\, 0.2,\, 0.4,\, 0.2]\ \si{\kilogram\square\meter}$ \\
		& $J_{xy}^{{\mathcal{L}}_{\bullet}^{\mathcal{R}_{\bullet}}}$ & first PoI of $\mathcal{L}_{\bullet}^{\mathcal{R}_{\bullet}}$ at the CoM of $\mathcal{L}_{\bullet}^{\mathcal{R}_{\bullet}}$ written in $\mathcal{R}_{l^{\mathcal{R}_{\bullet}}_{\bullet}}$ & $[0,\, 0,\, 0,\, 0,\, 0,\, 0]\ \si{\kilogram\square\meter}$ \\
		& $J_{xz}^{{\mathcal{L}}_{\bullet}^{\mathcal{R}_{\bullet}}}$ & second PoI of $\mathcal{L}_{\bullet}^{\mathcal{R}_{\bullet}}$ at the CoM of $\mathcal{L}_{\bullet}^{\mathcal{R}_{\bullet}}$ written in $\mathcal{R}_{l^{\mathcal{R}_{\bullet}}_{\bullet}}$ & $[0,\, 0,\, 0,\, 0,\, 0,\, 0]\ \si{\kilogram\square\meter}$ \\		
		& $J_{yz}^{{\mathcal{L}}_{\bullet}^{\mathcal{R}_{\bullet}}}$ & third PoI of $\mathcal{L}_{\bullet}^{\mathcal{R}_{\bullet}}$ at the CoM of $\mathcal{L}_{\bullet}^{\mathcal{R}_{\bullet}}$ written in $\mathcal{R}_{l^{\mathcal{R}_{\bullet}}_{\bullet}}$ & $[0,\, 0,\, 0,\, 0,\, 0,\, 0]\ \si{\kilogram\square\meter}$ \\	
  &
  $\mathbf{P}_{{\mathcal{R}_{{c}}}/{\mathcal{R}_{{l^{\mathcal{R}_{1}}_5}}}}$ & change of frame DCM between ${\mathcal{R}_{{c}}}=\left(J_{6}^{\mathcal{R}_{1/2}}; \mathbf{x}_{c}, \mathbf{y}_{c}, \mathbf{z}_{c}\right)$ and ${\mathcal{R}_{{l^{\mathcal{R}_{1}}_5}}}$ & $\mathbf{I}_3$ \\

    &
  $\mathbf{P}_{{\mathcal{R}_{{c}}}/{\mathcal{R}_{{l^{\mathcal{R}_{2}}_5}}}}$ & change of frame DCM between ${\mathcal{R}_{{c}}}$ and ${\mathcal{R}_{{l^{\mathcal{R}_{2}}_5}}}$ & $\begin{bmatrix}
		-0.5 & 0 & -0.866 \\
		 0 &  -1 & 0 \\
		-0.866 & 0 &  0.5 \\
		\end{bmatrix} $ \\

&
  $\mathbf{P}_{{\mathcal{R}_{{t}}}/{\mathcal{R}_{{l^{\mathcal{R}_{3}}_0}}}}$ & change of frame DCM between ${\mathcal{R}_{{t}}}=\left(J_{0}^{\mathcal{R}_{3}}; \mathbf{x}_{t}, \mathbf{y}_{t}, \mathbf{z}_{t}\right)$ and ${\mathcal{R}_{{l^{\mathcal{R}_{3}}_0}}}$ & $\mathbf{I}_3$ \\
  
  \midrule
		
		\multirow{16}{1.5cm}{\centering Multi-arm robot's central hub $\mathcal{\mathcal{C}}$}

  & ${\mathbf{DJ_6^{\mathcal{R}_1}}}$ & distance vector between $D$ and $J_6^{\mathcal{R}_1}$   written in $\mathcal{R}_{c} $  & $[0.1,\, 0,\, 0]$ \si{\meter} \\

    & ${\mathbf{DJ_6^{\mathcal{R}_2}}}$ & distance vector between  $D$ and $J_6^{\mathcal{R}_2}$  written in $\mathcal{R}_{c}$ & $[-0.05,\, 0,\, -0.0866]$ \si{\meter} \\

    & ${\mathbf{DJ_6^{\mathcal{R}_3}}}$ & distance vector between  $D$ and $J_6^{\mathcal{R}_3}$  written in $\mathcal{R}_{c}$ & $[-0.05,\, 0,\, 0.0866]$ \si{\meter} \\
		
		&  $m^{{\mathcal{C}}}$ & mass of $\mathcal{C}$ & $10\ \si{\kilogram}$ \\
		& $\begin{bmatrix}
			J_{xx}^{\mathcal{C}} & J_{xy}^{\mathcal{C}} & J_{xz}^{\mathcal{C}} \\
			 &  J_{yy}^{\mathcal{C}} & J_{yz}^{\mathcal{C}} \\
			 &  &  J_{zz}^{\mathcal{C}} \\
		\end{bmatrix}$ & inertia of $\mathcal{C}$ at the CoM $D$ written in $\mathcal{R}_{c}$ & $\begin{bmatrix}
		 0.6  & 0 & 0 \\
		 &  0.6  & 0 \\
		 &  &  0.6 \\
		\end{bmatrix}\, \si{\kilogram\square\meter} $ \\
  
  &
  $\mathbf{P}_{{\mathcal{R}_{{l^{\mathcal{R}_{1}}_5}}}/{\mathcal{R}_{c}}}$ & change of frame DCM between ${\mathcal{R}_{{l^{\mathcal{R}_{1}}_5}}}$ and ${\mathcal{R}_{c}}$ & $\mathbf{I}_3 $ \\

    &
  $\mathbf{P}_{{\mathcal{R}_{{l^{\mathcal{R}_{2}}_5}}}/{\mathcal{R}_{c}}}$ & change of frame DCM between ${\mathcal{R}_{{l^{\mathcal{R}_{2}}_5}}}$ and ${\mathcal{R}_{c}}$ & $\begin{bmatrix}
		-0.5 & 0 & -0.866 \\
		 0 &  -1 & 0 \\
		-0.866 & 0 &  0.5 \\
		\end{bmatrix} $ \\
&
  $\mathbf{P}_{{\mathcal{R}_{{l^{\mathcal{R}_{3}}_5}}}/{\mathcal{R}_{c}}}$ & change of frame DCM between ${\mathcal{R}_{{l^{\mathcal{R}_{3}}_5}}}$ and ${\mathcal{R}_{c}}$ & $\begin{bmatrix}
		-0.5 & 0 & 0.866 \\
		 0 &  -1 & 0 \\
		0.866 & 0 &  0.5 \\
		\end{bmatrix} $ \\
  \midrule
		
			\multirow{6}{1.5cm}{\centering Modular tile $\mathcal{T}$}
		
		&  $m^{{\mathcal{T}}}$ & mass of $\mathcal{T}$ & $6.0423\ \si{\kilogram}$ \\
		& $\begin{bmatrix}
			J_{xx}^{\mathcal{T}} & J_{xy}^{\mathcal{T}} & J_{xz}^{\mathcal{T}} \\
			 &  J_{yy}^{\mathcal{T}} & J_{yz}^{\mathcal{T}} \\
			 &  &  J_{zz}^{\mathcal{T}} \\
		\end{bmatrix}$ & inertia of $\mathcal{T}$ at $J_{0}^{\mathcal{R}_3}$ written in $\mathcal{R}_{t}$ & $\begin{bmatrix}
		 0.5041  & 0 & 0 \\
		 &  0.5041 & 0 \\
		 &  &  1.0071 \\
		\end{bmatrix}\, \si{\kilogram\square\meter} $ \\ 
		\\ \midrule

	\end{tabular}
	
}
\end{table}

\section*{Funding Sources}

This project received funding from ISAE-SUPAERO and the European Space Agency (ESA) through an Open Space Innovation Platform (OSIP) contract (grant number 4000143404).




\bibliographystyle{model1-num-names}
\bibliography{library.bib}

\begin{thebibliography}{43}
\expandafter\ifx\csname natexlab\endcsname\relax\def\natexlab#1{#1}\fi
\providecommand{\bibinfo}[2]{#2}
\ifx\xfnm\relax \def\xfnm[#1]{\unskip,\space#1}\fi
\bibitem[{Briz~Valero et~al.(2023)Briz~Valero, Paulino, Lourenço, Cachim, Forgues-Mayet, Groth, Papadopoulos, Rekleitis, Nanos, and Valentin}]{antennaGMV}
\bibinfo{author}{J.~Briz~Valero}, \bibinfo{author}{N.~Paulino}, \bibinfo{author}{P.~Lourenço}, \bibinfo{author}{P.~Cachim}, \bibinfo{author}{E.~Forgues-Mayet}, \bibinfo{author}{A.~Groth}, \bibinfo{author}{E.~Papadopoulos}, \bibinfo{author}{G.~Rekleitis}, \bibinfo{author}{K.~Nanos}, \bibinfo{author}{P.~Valentin},
\newblock \bibinfo{title}{Guidance, navigation, and control of in-orbit assembly of large antennas - technologies and approach for ioant}.
\bibitem[{Pirat et~al.(2022)Pirat, Ribes-Pleguezuelo, Keller, Zuccaro~Marchi, and Walker}]{pirat2022}
\bibinfo{author}{C.~Pirat}, \bibinfo{author}{P.~Ribes-Pleguezuelo}, \bibinfo{author}{F.~Keller}, \bibinfo{author}{A.~Zuccaro~Marchi}, \bibinfo{author}{R.~Walker},
\newblock \bibinfo{title}{Toward the autonomous assembly of large telescopes using cubesat rendezvous and docking},
\newblock \bibinfo{journal}{Journal of Spacecraft and Rockets} \bibinfo{volume}{59} (\bibinfo{year}{2022}) \bibinfo{pages}{375--388}.
\bibitem[{Santiago-Prowald and Baier(2013)}]{santiago2013}
\bibinfo{author}{J.~Santiago-Prowald}, \bibinfo{author}{H.~Baier},
\newblock \bibinfo{title}{Advances in deployable structures and surfaces for large apertures in space},
\newblock \bibinfo{journal}{CEAS Space Journal} \bibinfo{volume}{5} (\bibinfo{year}{2013}) \bibinfo{pages}{89--115}.
\bibitem[{Puig et~al.(2010)Puig, Barton, and Rando}]{PUIG201012}
\bibinfo{author}{L.~Puig}, \bibinfo{author}{A.~Barton}, \bibinfo{author}{N.~Rando},
\newblock \bibinfo{title}{A review on large deployable structures for astrophysics missions},
\newblock \bibinfo{journal}{Acta Astronautica} \bibinfo{volume}{67} (\bibinfo{year}{2010}) \bibinfo{pages}{12--26}.
\bibitem[{Ma et~al.(2022)Ma, Li, Ma, Wang, Shi, Zheng, Cui, Li, Liu, Guo, Liu, Wang, and Li}]{MA2022207}
\bibinfo{author}{X.~Ma}, \bibinfo{author}{T.~Li}, \bibinfo{author}{J.~Ma}, \bibinfo{author}{Z.~Wang}, \bibinfo{author}{C.~Shi}, \bibinfo{author}{S.~Zheng}, \bibinfo{author}{Q.~Cui}, \bibinfo{author}{X.~Li}, \bibinfo{author}{F.~Liu}, \bibinfo{author}{H.~Guo}, \bibinfo{author}{L.~Liu}, \bibinfo{author}{Z.~Wang}, \bibinfo{author}{Y.~Li},
\newblock \bibinfo{title}{Recent advances in space-deployable structures in china},
\newblock \bibinfo{journal}{Engineering} \bibinfo{volume}{17} (\bibinfo{year}{2022}) \bibinfo{pages}{207--219}.
\bibitem[{Yang et~al.(2023)Yang, Lu, and Xia}]{YANG2023108406}
\bibinfo{author}{C.~Yang}, \bibinfo{author}{W.~Lu}, \bibinfo{author}{Y.~Xia},
\newblock \bibinfo{title}{Uncertain optimal attitude control for space power satellite based on interval riccati equation with non-probabilistic time-dependent reliability},
\newblock \bibinfo{journal}{Aerospace Science and Technology} \bibinfo{volume}{139} (\bibinfo{year}{2023}) \bibinfo{pages}{108406}.
\bibitem[{Urbina et~al.(2023)Urbina, Prendergast, Madakashira, Roy, Barrio, and Deremetz}]{urbina2023}
\bibinfo{author}{D.~Urbina}, \bibinfo{author}{M.~Prendergast}, \bibinfo{author}{H.~Madakashira}, \bibinfo{author}{T.~Roy}, \bibinfo{author}{A.~Barrio}, \bibinfo{author}{M.~Deremetz},
\newblock \bibinfo{title}{Skybeam: In-orbit assembly for space-based solar power with european technologies}.
\bibitem[{Kulu(2023)}]{kulu2023}
\bibinfo{author}{E.~Kulu},
\newblock \bibinfo{title}{Space solar power - 2023 survey of public and private initiatives}.
\bibitem[{Wang et~al.(2021)Wang, Wu, Xun, Liu, and Wu}]{wang2021}
\bibinfo{author}{E.~Wang}, \bibinfo{author}{S.~Wu}, \bibinfo{author}{G.~Xun}, \bibinfo{author}{Y.~Liu}, \bibinfo{author}{Z.~Wu},
\newblock \bibinfo{title}{Active vibration suppression for large space structure assembly: A distributed adaptive model predictive control approach},
\newblock \bibinfo{journal}{Journal of Vibration and Control} \bibinfo{volume}{27} (\bibinfo{year}{2021}) \bibinfo{pages}{365--377}.
\bibitem[{She et~al.(2019)She, Li, and Wang}]{she2019}
\bibinfo{author}{Y.~She}, \bibinfo{author}{S.~Li}, \bibinfo{author}{Z.~Wang},
\newblock \bibinfo{title}{Constructing a large antenna reflector via spacecraft formation flying and reconfiguration control},
\newblock \bibinfo{journal}{Journal of Guidance, Control, and Dynamics} \bibinfo{volume}{42} (\bibinfo{year}{2019}) \bibinfo{pages}{1372--1382}.
\bibitem[{Nanjangud et~al.(2019)Nanjangud, Underwood, Bridges, Saaj, Eckersley, Sweeting, and Biancod}]{Nanjangud}
\bibinfo{author}{A.~Nanjangud}, \bibinfo{author}{C.~I. Underwood}, \bibinfo{author}{C.~P. Bridges}, \bibinfo{author}{C.~M. Saaj}, \bibinfo{author}{S.~Eckersley}, \bibinfo{author}{M.~N. Sweeting}, \bibinfo{author}{P.~Biancod},
\newblock \bibinfo{title}{Towards robotic on-orbit assembly of large space telescopes: Mission architectures, concepts, and analyses}.
\bibitem[{Lee et~al.(2016)Lee, Burdick, Backes, Pellegrino, Hogstrom, Fuller, Kennedy, Kim, Mukherjee, Seubert, and Wu}]{Lee}
\bibinfo{author}{N.~N. Lee}, \bibinfo{author}{J.~W. Burdick}, \bibinfo{author}{P.~Backes}, \bibinfo{author}{S.~Pellegrino}, \bibinfo{author}{K.~Hogstrom}, \bibinfo{author}{C.~Fuller}, \bibinfo{author}{B.~Kennedy}, \bibinfo{author}{J.~Kim}, \bibinfo{author}{R.~Mukherjee}, \bibinfo{author}{C.~Seubert}, \bibinfo{author}{Y.-H. Wu},
\newblock \bibinfo{title}{{Architecture for in-space robotic assembly of a modular space telescope}},
\newblock \bibinfo{journal}{Journal of Astronomical Telescopes, Instruments, and Systems} \bibinfo{volume}{2} (\bibinfo{year}{2016}) \bibinfo{pages}{041207}.
\bibitem[{Oegerle et~al.(2006)Oegerle, Purves, Budinoff, Moe, Carnahan, Evans, and Kim}]{Oegerle}
\bibinfo{author}{W.~R. Oegerle}, \bibinfo{author}{L.~R. Purves}, \bibinfo{author}{J.~G. Budinoff}, \bibinfo{author}{R.~V. Moe}, \bibinfo{author}{T.~M. Carnahan}, \bibinfo{author}{D.~C. Evans}, \bibinfo{author}{C.~K. Kim},
\newblock \bibinfo{title}{{Concept for a large scalable space telescope: in-space assembly}},
\newblock in: \bibinfo{editor}{J.~C. Mather}, \bibinfo{editor}{H.~A. MacEwen}, \bibinfo{editor}{M.~W.~M. de~Graauw} (Eds.), \bibinfo{booktitle}{Space Telescopes and Instrumentation I: Optical, Infrared, and Millimeter}, volume \bibinfo{volume}{6265}, \bibinfo{organization}{International Society for Optics and Photonics}, \bibinfo{publisher}{SPIE}, \bibinfo{year}{2006}, p. \bibinfo{pages}{62652C}.
\bibitem[{Jenett and Cheung(????)}]{jenett}
\bibinfo{author}{B.~Jenett}, \bibinfo{author}{K.~Cheung}, \bibinfo{title}{BILL-E: Robotic Platform for Locomotion and Manipulation of Lightweight Space Structures}.
\bibitem[{Letier et~al.(2020)Letier, Siedel, Deremetz, Pavlovskis, Lietaer, Nottensteiner, Roa, Sánchez García~Casarrubios, Romero, and Gancet}]{hotdock}
\bibinfo{author}{P.~Letier}, \bibinfo{author}{T.~Siedel}, \bibinfo{author}{M.~Deremetz}, \bibinfo{author}{E.~Pavlovskis}, \bibinfo{author}{B.~Lietaer}, \bibinfo{author}{K.~Nottensteiner}, \bibinfo{author}{M.~A. Roa}, \bibinfo{author}{J.~Sánchez García~Casarrubios}, \bibinfo{author}{J.~Romero}, \bibinfo{author}{J.~Gancet},
\newblock \bibinfo{title}{Hotdock: Design and validation of a new generation of standard robotic interface for on-orbit servicing}.
\bibitem[{Deremetz et~al.(2020)Deremetz, Letier, Grunwald, Roa, Brunner, and Lietaer}]{mosar}
\bibinfo{author}{M.~Deremetz}, \bibinfo{author}{P.~Letier}, \bibinfo{author}{G.~Grunwald}, \bibinfo{author}{M.~A. Roa}, \bibinfo{author}{B.~Brunner}, \bibinfo{author}{B.~Lietaer},
\newblock \bibinfo{title}{Mosar-wm: A relocatable robotic arm demonstrator for future on-orbit applications}.
\bibitem[{Deremetz et~al.(2022)Deremetz, Debroise, Govindaraj, But, Nieto, De~Stefano, Mishra, Brunner, Grunwald, Roa, Reiner, Závodník, Komarek, D’Amico, Cavenago, Gancet, Letier, Ilzkovitz, Gerdes, and Zwick}]{mirrormar}
\bibinfo{author}{M.~Deremetz}, \bibinfo{author}{M.~Debroise}, \bibinfo{author}{S.~Govindaraj}, \bibinfo{author}{A.~But}, \bibinfo{author}{I.~Nieto}, \bibinfo{author}{M.~De~Stefano}, \bibinfo{author}{H.~Mishra}, \bibinfo{author}{B.~Brunner}, \bibinfo{author}{G.~Grunwald}, \bibinfo{author}{M.~A. Roa}, \bibinfo{author}{M.~Reiner}, \bibinfo{author}{M.~Závodník}, \bibinfo{author}{M.~Komarek}, \bibinfo{author}{J.~D’Amico}, \bibinfo{author}{F.~Cavenago}, \bibinfo{author}{J.~Gancet}, \bibinfo{author}{P.~Letier}, \bibinfo{author}{M.~Ilzkovitz}, \bibinfo{author}{L.~Gerdes}, \bibinfo{author}{M.~Zwick},
\newblock \bibinfo{title}{Demonstrator design of a modular multi-arm robot for on-orbit large telescope assembly}.
\bibitem[{Wang et~al.(2019)Wang, Wu, Wu, and Radice}]{WANG2019105378}
\bibinfo{author}{E.~Wang}, \bibinfo{author}{S.~Wu}, \bibinfo{author}{Z.~Wu}, \bibinfo{author}{G.~Radice},
\newblock \bibinfo{title}{Distributed adaptive vibration control for solar power satellite during on-orbit assembly},
\newblock \bibinfo{journal}{Aerospace Science and Technology} \bibinfo{volume}{94} (\bibinfo{year}{2019}) \bibinfo{pages}{105378}.
\bibitem[{Dong et~al.(2024)Dong, Li, Ning, and Wang}]{DONG2024108959}
\bibinfo{author}{H.~Dong}, \bibinfo{author}{T.~Li}, \bibinfo{author}{Y.~Ning}, \bibinfo{author}{Z.~Wang},
\newblock \bibinfo{title}{Dynamic modeling and attitude control for large modular antennas on-orbit assembly},
\newblock \bibinfo{journal}{Aerospace Science and Technology} \bibinfo{volume}{146} (\bibinfo{year}{2024}) \bibinfo{pages}{108959}.
\bibitem[{Zhang et~al.(2024)Zhang, Sai, Li, Sun, Zhang, and Xu}]{zhang2024}
\bibinfo{author}{E.~Zhang}, \bibinfo{author}{H.~Sai}, \bibinfo{author}{Y.~Li}, \bibinfo{author}{X.~Sun}, \bibinfo{author}{T.~Zhang}, \bibinfo{author}{Z.~Xu},
\newblock \bibinfo{title}{Modular robotic manipulator and ground assembly system for on-orbit assembly of space telescopes},
\newblock \bibinfo{journal}{Proceedings of the Institution of Mechanical Engineers, Part C: Journal of Mechanical Engineering Science} \bibinfo{volume}{238} (\bibinfo{year}{2024}) \bibinfo{pages}{2283--2293}.
\bibitem[{Zhou et~al.(2023)Zhou, Wu, and Yang}]{Zhou2023}
\bibinfo{author}{W.~Zhou}, \bibinfo{author}{S.~Wu}, \bibinfo{author}{J.~Yang},
\newblock \bibinfo{title}{Modular dynamic modeling for on-orbit assembly of large-scale space structures},
\newblock \bibinfo{journal}{International Journal of Aerospace Engineering} \bibinfo{volume}{2023} (\bibinfo{year}{2023}) \bibinfo{pages}{6659124}.
\bibitem[{Swei et~al.(????)Swei, Jenett, Cramer, and Cheung}]{Sean2020}
\bibinfo{author}{S.~S.-M. Swei}, \bibinfo{author}{B.~Jenett}, \bibinfo{author}{N.~B. Cramer}, \bibinfo{author}{K.~Cheung}, \bibinfo{title}{Modeling and Control of Robot-Structure Coupling During In-Space Structure Assembly}.
\bibitem[{Gong and Macdonald(2019)}]{Gong2019}
\bibinfo{author}{S.~Gong}, \bibinfo{author}{M.~Macdonald},
\newblock \bibinfo{title}{Review on solar sail technology},
\newblock \bibinfo{journal}{Astrodynamics} \bibinfo{volume}{3} (\bibinfo{year}{2019}) \bibinfo{pages}{93--125}.
\bibitem[{Boning and Dubowsky(2010)}]{Peggy}
\bibinfo{author}{P.~Boning}, \bibinfo{author}{S.~Dubowsky},
\newblock \bibinfo{title}{Coordinated control of space robot teams for the on-orbit construction of large flexible space structures},
\newblock \bibinfo{journal}{Advanced Robotics} \bibinfo{volume}{24} (\bibinfo{year}{2010}) \bibinfo{pages}{303--323}.
\bibitem[{Yoshida(2000)}]{yoshida}
\bibinfo{author}{K.~Yoshida},
\newblock \bibinfo{title}{Space robot dynamics and control: To orbit, from orbit, and future},
\newblock in: \bibinfo{editor}{J.~M. Hollerbach}, \bibinfo{editor}{D.~E. Koditschek} (Eds.), \bibinfo{booktitle}{Robotics Research}, \bibinfo{publisher}{Springer London}, \bibinfo{address}{London}, \bibinfo{year}{2000}, pp. \bibinfo{pages}{449--456}.
\bibitem[{Maghami and Lim(2009)}]{maghami}
\bibinfo{author}{P.~Maghami}, \bibinfo{author}{K.~Lim},
\newblock \bibinfo{title}{Synthesis and control of flexible systems with component-level uncertainties},
\newblock \bibinfo{journal}{Journal of Dynamic Systems Measurement and Control-transactions of The Asme - J DYN SYST MEAS CONTR} \bibinfo{volume}{131} (\bibinfo{year}{2009}).
\bibitem[{Alazard et~al.(????)Alazard, Perez, Cumer, and Loquen}]{alazard}
\bibinfo{author}{D.~Alazard}, \bibinfo{author}{J.~A. Perez}, \bibinfo{author}{C.~Cumer}, \bibinfo{author}{T.~Loquen}, \bibinfo{title}{Two-input two-output port model for mechanical systems}.
\bibitem[{Murali et~al.(2015)Murali, Alazard, Massotti, Ankersen, and Toglia}]{MURALI}
\bibinfo{author}{H.~H.~S. Murali}, \bibinfo{author}{D.~Alazard}, \bibinfo{author}{L.~Massotti}, \bibinfo{author}{F.~Ankersen}, \bibinfo{author}{C.~Toglia},
\newblock \bibinfo{title}{Mechanical-attitude controller co-design of large flexible space structures},
\newblock in: \bibinfo{editor}{J.~Bordeneuve-Guib{\'e}}, \bibinfo{editor}{A.~Drouin}, \bibinfo{editor}{C.~Roos} (Eds.), \bibinfo{booktitle}{Advances in Aerospace Guidance, Navigation and Control}, \bibinfo{publisher}{Springer International Publishing}, \bibinfo{address}{Cham}, \bibinfo{year}{2015}, pp. \bibinfo{pages}{659--678}.
\bibitem[{Perez et~al.(2015)Perez, Alazard, Loquen, Cumer, and Pittet}]{Perez2015}
\bibinfo{author}{J.~A. Perez}, \bibinfo{author}{D.~Alazard}, \bibinfo{author}{T.~Loquen}, \bibinfo{author}{C.~Cumer}, \bibinfo{author}{C.~Pittet},
\newblock \bibinfo{title}{Linear dynamic modeling of spacecraft with open-chain assembly of flexible bodies for {ACS}/structure co-design},
\newblock in: \bibinfo{booktitle}{Advances in Aerospace Guidance, Navigation and Control}, \bibinfo{publisher}{Springer International Publishing}, \bibinfo{year}{2015}, pp. \bibinfo{pages}{639--658}.
\bibitem[{Perez et~al.(2016)Perez, Alazard, Loquen, Pittet, and Cumer}]{Perez2016}
\bibinfo{author}{J.~A. Perez}, \bibinfo{author}{D.~Alazard}, \bibinfo{author}{T.~Loquen}, \bibinfo{author}{C.~Pittet}, \bibinfo{author}{C.~Cumer},
\newblock \bibinfo{title}{Flexible multibody system linear modeling for control using component modes synthesis and double-port approach},
\newblock \bibinfo{journal}{Journal of Dynamic Systems, Measurement, and Control} \bibinfo{volume}{138} (\bibinfo{year}{2016}).
\bibitem[{Sanfedino et~al.(2018)Sanfedino, Alazard, Pommier-Budinger, Falcoz, and Boquet}]{SANFEDINO2018128}
\bibinfo{author}{F.~Sanfedino}, \bibinfo{author}{D.~Alazard}, \bibinfo{author}{V.~Pommier-Budinger}, \bibinfo{author}{A.~Falcoz}, \bibinfo{author}{F.~Boquet},
\newblock \bibinfo{title}{Finite element based n-port model for preliminary design of multibody systems},
\newblock \bibinfo{journal}{Journal of Sound and Vibration} \bibinfo{volume}{415} (\bibinfo{year}{2018}) \bibinfo{pages}{128--146}.
\bibitem[{Rodrigues et~al.(2022)Rodrigues, Preda, Sanfedino, and Alazard}]{RODRIGUES2022107865}
\bibinfo{author}{R.~Rodrigues}, \bibinfo{author}{V.~Preda}, \bibinfo{author}{F.~Sanfedino}, \bibinfo{author}{D.~Alazard},
\newblock \bibinfo{title}{Modeling, robust control synthesis and worst-case analysis for an on-orbit servicing mission with large flexible spacecraft},
\newblock \bibinfo{journal}{Aerospace Science and Technology} \bibinfo{volume}{129} (\bibinfo{year}{2022}) \bibinfo{pages}{107865}.
\bibitem[{Rodrigues et~al.(2023)Rodrigues, Sanfedino, Alazard, Preda, and Olucha}]{rodrigues}
\bibinfo{author}{R.~Rodrigues}, \bibinfo{author}{F.~Sanfedino}, \bibinfo{author}{D.~Alazard}, \bibinfo{author}{V.~Preda}, \bibinfo{author}{J.~Olucha},
\newblock \bibinfo{title}{{Linear parameter-varying gain-scheduled attitude controller for an on-orbit servicing mission involving flexible large spacecraft and fuel sloshing}},
\newblock in: \bibinfo{booktitle}{{ESA GNC and ICATT Conference 2023}}, \bibinfo{address}{Sopot, Poland}.
\bibitem[{Rodrigues et~al.(2024)Rodrigues, Alazard, Sanfedino, Mauriello, and Iannelli}]{Rodrigues2024ModelingAA}
\bibinfo{author}{R.~Rodrigues}, \bibinfo{author}{D.~Alazard}, \bibinfo{author}{F.~Sanfedino}, \bibinfo{author}{T.~Mauriello}, \bibinfo{author}{P.~Iannelli},
\newblock \bibinfo{title}{Modeling and analysis of a flexible spinning euler-bernoulli beam with centrifugal stiffening and softening: A linear fractional representation approach with application to spinning spacecraft},
\newblock \bibinfo{journal}{ArXiv} \bibinfo{volume}{abs/2401.17519} (\bibinfo{year}{2024}).
\bibitem[{Alazard and Sanfedino(2021)}]{userguide}
\bibinfo{author}{D.~Alazard}, \bibinfo{author}{F.~Sanfedino}, \bibinfo{title}{Satellite dynamics toolbox library (sdtlib) - user's guide}, \bibinfo{year}{2021}.
\bibitem[{Chebbi et~al.(2017)Chebbi, Dubanchet, Perez~Gonzalez, and Alazard}]{CHEBBI}
\bibinfo{author}{J.~Chebbi}, \bibinfo{author}{V.~Dubanchet}, \bibinfo{author}{J.~A. Perez~Gonzalez}, \bibinfo{author}{D.~Alazard},
\newblock \bibinfo{title}{Linear dynamics of flexible multibody systems: A system-based approach},
\newblock \bibinfo{journal}{Multibody System Dynamics} \bibinfo{volume}{41} (\bibinfo{year}{2017}).
\bibitem[{Alazard et~al.(2023)Alazard, Finozzi, and Sanfedino}]{finozzi}
\bibinfo{author}{D.~Alazard}, \bibinfo{author}{A.~Finozzi}, \bibinfo{author}{F.~Sanfedino},
\newblock \bibinfo{title}{Port inversions of parametric two-input two-output port models of flexible substructures},
\newblock \bibinfo{journal}{Multibody System Dynamics} \bibinfo{volume}{57} (\bibinfo{year}{2023}).
\bibitem[{Guy et~al.(2014)Guy, Alazard, Cumer, and Charbonnel}]{Guy2014}
\bibinfo{author}{N.~Guy}, \bibinfo{author}{D.~Alazard}, \bibinfo{author}{C.~Cumer}, \bibinfo{author}{C.~Charbonnel},
\newblock \bibinfo{title}{Dynamic modeling and analysis of spacecraft with variable tilt of flexible appendages},
\newblock \bibinfo{journal}{Journal of Dynamic Systems, Measurement, and Control} \bibinfo{volume}{136} (\bibinfo{year}{2014}).
\bibitem[{Williamson(2010)}]{williamsonlists}
\bibinfo{author}{E.~Williamson}, \bibinfo{title}{Lists, Decisions and Graphs}, \bibinfo{publisher}{S. Gill Williamson}, \bibinfo{year}{2010}.
\bibitem[{Cormen et~al.(2009)Cormen, Leiserson, Rivest, and Stein}]{cormen}
\bibinfo{author}{T.~Cormen}, \bibinfo{author}{C.~Leiserson}, \bibinfo{author}{R.~Rivest}, \bibinfo{author}{C.~Stein}, \bibinfo{title}{Introduction to Algorithms, Third Edition}, \bibinfo{year}{2009}.
\bibitem[{Packard and Doyle(1993)}]{Doyle1993}
\bibinfo{author}{A.~Packard}, \bibinfo{author}{J.~Doyle},
\newblock \bibinfo{title}{{The complex structured singular value}},
\newblock \bibinfo{journal}{Automatica} \bibinfo{volume}{29} (\bibinfo{year}{1993}) \bibinfo{pages}{71--109}.
\bibitem[{Preda et~al.(2020)Preda, Sanfedino, Bennani, Boquet, and Alazard}]{Preda2020}
\bibinfo{author}{V.~Preda}, \bibinfo{author}{F.~Sanfedino}, \bibinfo{author}{S.~Bennani}, \bibinfo{author}{F.~Boquet}, \bibinfo{author}{D.~Alazard},
\newblock \bibinfo{title}{Robust and adaptable dynamic response reshaping of flexible structures},
\newblock \bibinfo{journal}{Journal of Sound and Vibration} \bibinfo{volume}{468} (\bibinfo{year}{2020}) \bibinfo{pages}{115086}.
\bibitem[{Balas et~al.(2007)Balas, Chiang, Packard, and Safonov}]{balas2007robust}
\bibinfo{author}{G.~Balas}, \bibinfo{author}{R.~Chiang}, \bibinfo{author}{A.~Packard}, \bibinfo{author}{M.~Safonov},
\newblock \bibinfo{title}{Robust control toolbox user’s guide},
\newblock \bibinfo{journal}{The Math Works, Inc., Tech. Rep}  (\bibinfo{year}{2007}).

\end{thebibliography}







\end{document}